\documentclass[aps,pra,superscriptaddress,10pt,twocolumn]{revtex4-2}
\usepackage{amsmath,amssymb,amsfonts}
\usepackage{wrapfig}
\usepackage{comment}
\usepackage{url}
\usepackage{ascmac}
\usepackage{amsthm}
\usepackage{mathrsfs}
\usepackage{siunitx}
\usepackage{hyperref}
\usepackage{mathtools}

\usepackage[utf8]{inputenc}
\usepackage{color}

\usepackage{braket}
\usepackage{bbold}
\usepackage{graphicx}
\usepackage[normalem]{ulem}
\usepackage{bm}
\hypersetup{colorlinks=true,linkcolor=blue,citecolor=blue,urlcolor=blue}
\newcommand{\C}{\mathbb{C}}
\newcommand{\R}{\mathbb{R}}

\begin{document}


\title{Quantum signal processing in Hilbert space fragmented systems}


\author{Naoya Egawa}
\affiliation{Department of Physics, Tohoku University, Sendai, Miyagi 980-8578, Japan}
\email{egawa.naoya.p4@dc.tohoku.ac.jp}

\author{Kaoru Mizuta}
\affiliation{Department of Applied Physics, Graduate School of Engineering, The University of Tokyo, Hongo 7-3-1, Bunkyo, Tokyo 113-8656, Japan}
\affiliation{Photon Science Center, Graduate School of Engineering, The University of Tokyo, Hongo 7-3-1, Bunkyo, Tokyo 113-8656, Japan}
\affiliation{RIKEN Center for Quantum Computing (RQC), Hirosawa 2-1, Wako, Saitama 351-0198, Japan}

\author{Joji Nasu}
\affiliation{Department of Physics, Tohoku University, Sendai, Miyagi 980-8578, Japan}


\date{\today}

\begin{abstract}


Quantum signal processing (QSP), originally developed for composite pulse sequences in nuclear magnetic resonance systems, has recently attracted attention as a unified framework for quantum algorithms. 
A pioneering study applied QSP to nonequilibrium control in integrable many-body systems, enabling the realization of nonequilibrium dynamics with greater flexibility than Floquet engineering. 
However, extending QSP to nonintegrable systems faces fundamental obstacles arising from the limited number of conserved quantities and thermalization. 
In this work, we propose a protocol that leverages QSP in systems exhibiting Hilbert space fragmentation (HSF). 
Specifically, we consider a pair-hopping model with four-fold periodic potentials that exhibits an HSF structure, thereby providing integrable and nonintegrable sectors within a single system. 
We analytically show that nonequilibrium dynamics can be flexibly designed through QSP engineered by these potentials in the integrable sectors. 
In contrast, we numerically identify signatures of thermalization in the nonintegrable sectors. 
Remarkably, by inserting domain walls, we achieve parallel control of multiple quantum dynamics within a single system.
This approach sheds light on the control of nonequilibrium dynamics from the perspective of quantum computation by extending the scope of QSP to nonintegrable systems.

\end{abstract}

\maketitle

\section{Introduction}
\begin{figure*}
        \includegraphics[width=7in]{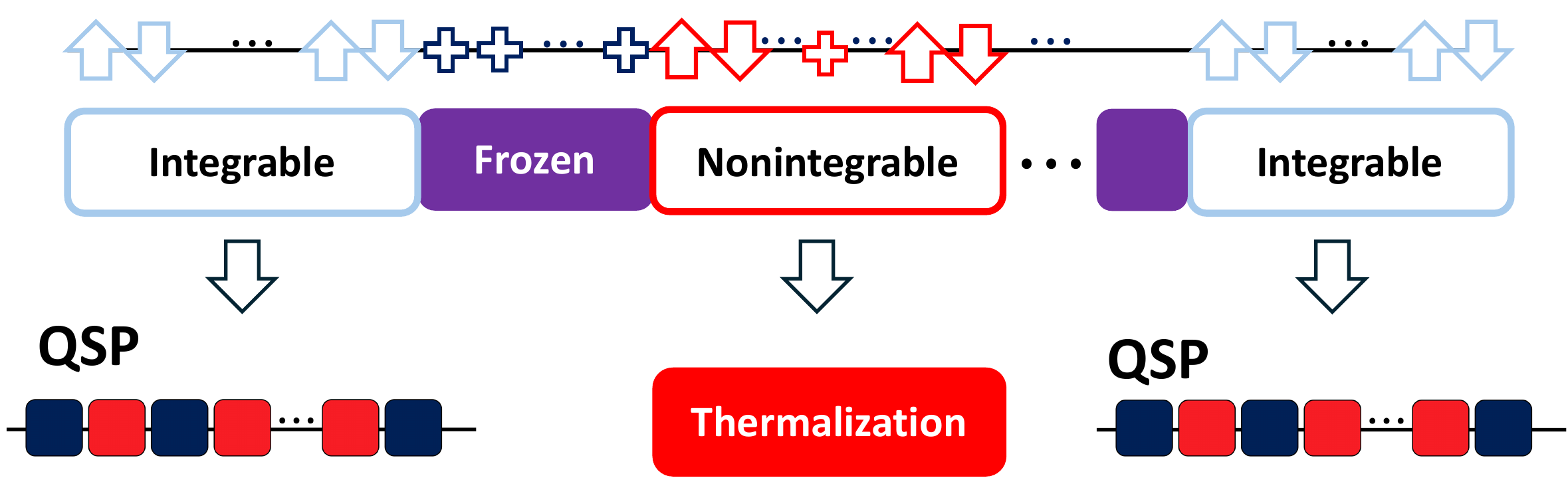}
    \caption{Schematic illustration of our proposal. 
    Depending on the initial pseudospin configurations shown at the top, the whole system is divided into integrable sectors (white boxes with light-blue outlines), nonintegrable sectors (white boxes with red outlines), and frozen regions (purple rectangles). In this paper, we analytically show that nonequilibrium dynamics can be independently controlled through QSP in the integrable sectors. In contrast, we also demonstrate signatures of thermalization in the nonintegrable sectors. }
    \label{fig:hsfqsp_all}
\end{figure*}
Designing nonequilibrium quantum dynamics in artificial quantum systems is of central interest in condensed matter physics. 
A practical challenge is that isolated quantum many-body systems generally reach thermal equilibrium under unitary dynamics, making it difficult to realize nonthermal behavior~\cite{deutschQuantumStatisticalMechanics1991,srednickiChaosQuantumThermalization1994,rigolThermalizationItsMechanism2008,rigolBreakdownThermalizationFinite2009,dalessioQuantumChaosEigenstate2016,moriThermalizationPrethermalizationIsolated2018, sugimotoBoundsEigenstateThermalization2025}.
To evade thermalization, the ability to control nonequilibrium dynamics in a flexible manner is therefore essential. 
In this respect, highly controllable quantum simulators provide a promising platform for a broad range of nonequilibrium phenomena,
 including many-body localization (MBL)~\cite{oganesyanLocalizationInteractingFermions2007,palManybodyLocalizationPhase2010,serbynLocalConservationLaws2013a,husePhenomenologyFullyManybodylocalized2014,luitzManybodyLocalizationEdge2015,nandkishoreManyBodyLocalizationThermalization2015,dalessioQuantumChaosEigenstate2016,abaninColloquiumManybodyLocalization2019,sierantManyBodyLocalizationAge2024}, prethermalization~\cite{kollarGeneralizedGibbsEnsemble2011, moriThermalizationPrethermalizationIsolated2018}, and quantum many-body scars (QMBS)~\cite{bernienProbingManybodyDynamics2017,shiraishiSystematicConstructionCounterexamples2017,turnerWeakErgodicityBreaking2018,moudgalyaThermalizationItsAbsence2020,markUnifiedStructureExact2020,serbynQuantumManybodyScars2021,papicWeakErgodicityBreaking2021,moudgalyaQuantumManyBodyScars2022,chandranQuantumManyBodyScars2023}. Likewise, as exemplified by Floquet engineering, discrete time crystals (DTCs)~\cite{elseFloquetTimeCrystals2016,yaoDiscreteTimeCrystals2017,khemaniPhaseStructureDriven2016,ippolitiManyBodyPhysicsNISQ2021,randallManybodylocalizedDiscrete2021,miTimecrystallineEigenstateOrder2022} and Floquet prethermalization~\cite{kollarGeneralizedGibbsEnsemble2011, moriThermalizationPrethermalizationIsolated2018,abaninRigorousTheoryManyBody2017, moriRigorousBoundsHeating2021,kuwaharaFloquetMagnusTheory2016} have been established as standard engineering strategies in periodically driven systems. However, these approaches often impose constraints on protocol design owing to the requirements of periodicity and drive frequency. Therefore, it is desirable to develop driving techniques that are not bound by such constraints. 
Our study aims to advance this standard engineering paradigm by exploiting quantum emulation~\cite{marvianEfficientQuantumEmulation2025}, namely, the idea of realizing the input--output relation of quantum dynamics. 
Note that quantum emulation is distinct from the concept of quantum simulation~\cite{feynmanSimulatingPhysicsComputers1982,lloydUniversalQuantumSimulators1996, georgescuQuantumSimulation2014}, which aims to realize the target dynamics step by step in time. 

In this context, quantum signal processing (QSP) has recently emerged as a systematic approach to quantum emulation and offers a systematic alternative to Floquet engineering~\cite{bastidasQuantumSignalProcessing2024}. At its core, QSP realizes a desired single-qubit dynamics by interleaving repeated applications of a signal-dependent unitary with controllable signal-processing operations, so that the resulting transformation exhibits a prescribed polynomial dependence on the underlying signal~\cite{lowMethodologyResonantEquiangular2016}. 
The broad expressivity of QSP has further motivated the formulation of quantum singular value transformation (QSVT)~\cite{martynGrandUnificationQuantum2021, gilyenQuantumSingularValue2019}, which applies polynomial transformations to the singular values of a block-encoded linear operator. QSVT is widely regarded as the ``grand unification'' of quantum algorithms, as it provides a unified description of various quantum algorithms such as Hamiltonian simulation and quantum phase estimation~\cite{groverFastQuantumMechanical1996,brassardQuantumAmplitudeAmplification2000,rallAmplitudeEstimationQuantum2023,lowOptimalHamiltonianSimulation2017,lowHamiltonianSimulationQubitization2019}. 

Building on these observations, Bastidas \textit{et al.}~\cite{bastidasQuantumSignalProcessing2024} employed a QSP-based emulation technique in the one-dimensional transverse-field Ising model (1D TFIM), taking momentum as the signal. 
While this approach successfully emulates nonequilibrium dynamics without requiring nonlocal or conditional operations that require additional ancilla resources, it relies on the integrability of the model. 
In contrast, generic quantum simulators typically exhibit chaotic behavior due to the nonintegrability of the Hamiltonian, and their unitary dynamics thermalize, accompanied by a local loss of information. 

In this paper, we explore the QSP protocol in a nonintegrable quantum many-body system. 
For generic nonintegrable systems with at most a few global conserved quantities, it is not straightforward to map the dynamics onto the Bogoliubov--de Gennes (BdG) structure exploited in prior QSP-based emulation schemes. 
Moreover, nonintegrability typically leads to thermalization, which is incompatible with sustaining nonequilibrium dynamics. 
Our goal is to demonstrate that nonequilibrium dynamics can nevertheless be emulated in an exceptional class of nonintegrable systems and to elucidate how extending beyond integrability affects controllability and its limitations relative to prior QSP constructions. 
To this end, we focus on a system exhibiting Hilbert space fragmentation (HSF), where the Hilbert space decomposes into dynamically disconnected subsectors and exhibits signatures of ergodic or nonergodic behavior depending on the initial state~\cite{salaErgodicityBreakingArising2020,rakovszkyStatisticalLocalizationStrong2020,yoshinagaEmergenceHilbertSpace2022,moudgalyaQuantumManyBodyScars2022a,khemaniLocalizationHilbertSpace2020,moudgalyaThermalizationItsAbsence2020,moudgalyaQuantumManybodyScars2020,chandranQuantumManyBodyScars2023}. We demonstrate that the target unitary evolution generated by a variety of BdG-integrable Hamiltonians can be implemented via QSP within an integrable sector of a whole nonintegrable system. 
Such integrable sectors can be generated in multiple copies by partitioning the system with domain walls. 
This enables parallel QSP-based emulation across spatial regions belonging to different integrable sectors, all within a single system [Fig.~\ref{fig:hsfqsp_all}]. 
Furthermore, we clarify that our protocols exhibit both thermal and nonthermal properties across subsectors.  
These results extend the conventional QSP strategy by exploiting the HSF structure, with potential applications to programmable nonequilibrium control in cold-atom quantum simulators.

This paper is organized as follows. In Sec.~\ref{sec2}, we provide a brief overview of QSP. In Sec.~\ref{sec3}, we introduce a representative model exhibiting HSF, which serves as a platform for QSP in a nonintegrable system. In Sec.~\ref{sec4}, we present our main results on QSP in HSF systems.  
Finally, Sec.~\ref{sec5} is devoted to the conclusion and outlook.

\section{Overview: Quantum signal processing}~\label{sec2}
\begin{figure*}
\includegraphics[width=7in]{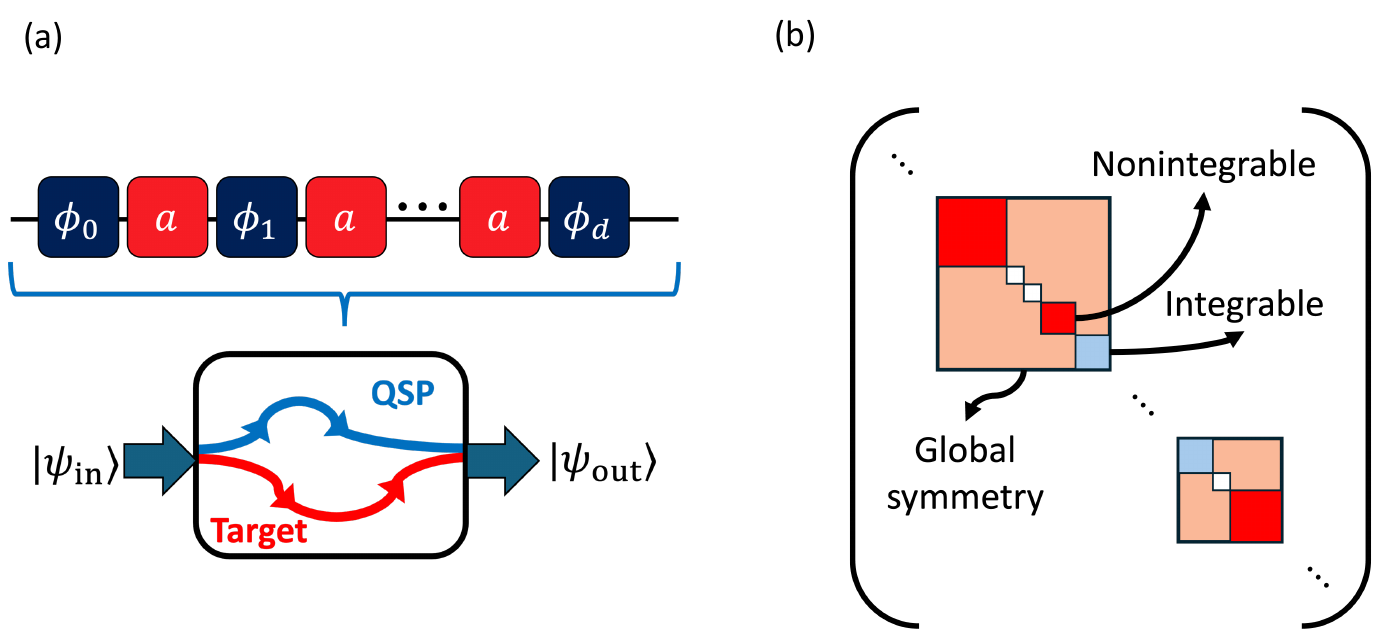}
\caption{(a) Illustration of QSP. In the gate structure, the orange blocks labeled by $a$ denote the signal operators and the navy blocks labeled by $\vec\phi=(\phi_0,\phi_1,\dots,\phi_d)$ denote the signal-processing operators. The QSP sequence (blue curve) emulates the target unitary dynamics (red curve) for a given input state $\ket{\psi_{\rm in}}$ and output state $\ket{\psi_{\rm out}}$. 
(b) Schematic picture of Hilbert space fragmentation. In addition to the global symmetries of the Hamiltonian, kinetic constraints cause the Hilbert space to decompose into a direct sum of Krylov subspaces. The dimension of some fragments can grow with the increase in system size. The Hamiltonian restricted to each fragment can be integrable (blue block) or nonintegrable (red block). 
}
\label{fig:qsp}
\end{figure*}

We briefly review QSP, a quantum emulation technique that enables polynomial transformations of a signal encoded in a unitary operator~\cite{lowMethodologyResonantEquiangular2016}. 
We begin with the signal operator, parameterized by an uncontrollable signal parameter $a$: 
\begin{equation}
\label{eq:signal}
W(a) = \exp(-iX\theta /2) = 
\begin{pmatrix}
    a & i\sqrt{1-a^{2}} \\
    i\sqrt{1-a^{2}} & a 
    \end{pmatrix}, 
\end{equation}
where $X$ denotes the Pauli-$X$ operator, and $\theta = -2\cos^{-1}a$ ($a \in [-1,+1]$) is the rotation angle about the $x$-axis on the Bloch sphere. 
To generate polynomial transformations of the signal $a$, we introduce a signal-processing operator that implements a rotation about an axis orthogonal to that of $W(a)$, 
\begin{align}
S(\phi) = \exp(iZ\phi) = 
\begin{pmatrix}
    e^{i\phi} & 0 \\
    0 & e^{-i\phi} \\
    \end{pmatrix}, 
\label{eq:signalprocessing}
\end{align}
where $Z$ denotes the Pauli-$Z$ operator, and $\phi$ is a controllable parameter specifying a rotation about the $z$-axis by an angle $-2\phi$. 
By combining these operations, we obtain the $(W_x, S_z)$-QSP sequence 
\begin{align}
    U_{\vec{\phi}}(a) = S(\phi_{0}) \prod_{r=1}^{d} \bigl[W(a) S(\phi_{r})\bigr], 
    \label{eq:qsp-single}
\end{align}
with a phase vector $\vec\phi=(\phi_0,\phi_1,\dots,\phi_d)\in \R^{\,d+1}$. 
A central result in QSP characterizes the set of unitary operators achievable through control of $\vec{\phi}$~\cite{lowHamiltonianSimulationQubitization2019,gilyenQuantumSingularValue2019, martynGrandUnificationQuantum2021,martynEfficientFullycoherentQuantum2023,yamamotoRobustAngleFinding2024, laneveMultivariatePolynomialsAchievable2025a}. 
Specifically, we consider a unitary operator parameterized by complex polynomials $P(a), Q(a) \in \C[a]$: 
\begin{align}
    U_{PQ}(a)  =
    \begin{pmatrix}
        P(a) & i Q(a) \sqrt{1 - a^2} \\
        i Q^*(a) \sqrt{1 - a^2} & P^*(a)
      \end{pmatrix}. 
      \label{eq:qsp_poly}
\end{align}
The theorem states that $\{U_{PQ}(a)\}_{P, Q} = \{U_{\vec{\phi}}(a)\}_{\vec{\phi}}$ if and only if the following conditions hold: 
    \begin{description}
        \item[1. Degree] 
            \begin{equation}
                \deg(P) \le d, \quad \deg(Q) \le d-1,\label{eq:degree}
            \end{equation}
        \item[2. Parity] 
            \begin{equation}
                P(-a) = (-1)^{d}P(a), \quad Q(-a) = (-1)^{d-1}Q(a), 
            \end{equation}
        \item[3. Unitarity] 
             \begin{equation}
                 |P(a)|^{2} + (1-a^{2})|Q(a)|^{2} = 1.\label{eq:unitarity}
             \end{equation}
    \end{description}
Consequently, for any $P(a)$ and $Q(a)$ satisfying the above conditions, there exists a phase sequence $\vec{\phi}$ that realizes $U_{PQ}(a)$, and conversely. 
Thus, QSP enables the implementation of a prescribed degree-$d$ polynomial transformation of the signal $a$ through the design of a phase sequence $\vec{\phi}$ of length $d+1$ [Fig.~\ref{fig:qsp}(a)]. 

\section{Overview: Hilbert space fragmentation}~\label{sec3}
We present a concrete model exhibiting HSF that serves as a platform for applying the aforementioned QSP in the subsequent sections. 
Previous studies have shown that HSF emerges in a variety of many-body systems~\cite{moudgalyaThermalizationItsAbsence2020,moudgalyaQuantumManybodyScars2020,salaErgodicityBreakingArising2020,rakovszkyStatisticalLocalizationStrong2020,yoshinagaEmergenceHilbertSpace2022,moudgalyaQuantumManyBodyScars2022a}. 
In this paper, motivated by our aim of studying a simple low-dimensional setting, we consider the pair-hopping model on a one-dimensional chain of spinless fermions. 
The Hamiltonian of this model is given by~\cite{moudgalyaThermalizationItsAbsence2020, moudgalyaQuantumManybodyScars2020,moudgalyaQuantumManyBodyScars2022a}
\begin{align}
    H_{\rm PH} = J\sum\limits_{j=1}^{L-3} (c_{j}^{\dagger}c_{j+3}^{\dagger}c_{j+2}c_{j+1} + h.c.), 
    \label{eq:pair-hopping}
\end{align}
where $J$ is the coupling constant, $c_j^{\dagger}$ and $c_j$ are the creation and annihilation operators for a spinless fermion at site $j$, respectively, and $L$ is the number of sites. 
For experimental feasibility, we impose open boundary conditions (OBC). 
In cold-atom realizations with a strong electric field, Eq.~\eqref{eq:pair-hopping} describes the lowest-order effective hopping term in the Stark MBL regime~\cite{moudgalyaQuantumManybodyScars2020, schulzStarkManyBodyLocalization2019, vannieuwenburgBlochOscillationsManybody2019,boidiCrossoverWannierStarkLocalization2025}. 
The full Hamiltonian $H_{\rm PH}$ has a limited number of global symmetries, such as the total particle number $N_{\rm tot}$ and the center-of-mass position $C$:
\begin{align}
    N_{\rm tot} = &\sum_{j=1}^{L} n_{j}, \label{eq:particlenumberconservation}\\
    C = &\sum_{j=1}^{L} jn_{j}\label{eq:comconservation}, 
\end{align}
where $n_j = c_j^{\dagger}c_j$. In addition, for even $L$, $H_{\rm PH}$ conserves the sublattice particle numbers: 
\begin{align}
    N_{e} = \sum_{j=1}^{L/2} n_{2j},\ N_{o} = \sum_{j=1}^{L/2} n_{2j-1}. 
\end{align}
In the simplest $4$-site setting, the Hamiltonian acts nontrivially within the subspace spanned by the configurations $\ket{1001} \rightleftarrows \ket{0110}$, where $\ket{abcd}$ denotes the fermionic occupation configuration on four consecutive sites. That is, two fermions are squeezed and anti-squeezed through these transitions~\cite{moudgalyaQuantumManybodyScars2020}. 
In contrast, the Hamiltonian annihilates all other configurations under its action, creating frozen regions. 
These conservation laws, together with the restricted local action of Eq.~\eqref{eq:pair-hopping}, give rise to kinetic constraints~\cite{salaErgodicityBreakingArising2020,khemaniLocalizationHilbertSpace2020}.
As a result of these global symmetries and kinetic constraints, the Hilbert space splits into an exponential number of Krylov subspaces [Fig.~\ref{fig:qsp}(b)]. 
This phenomenon is known as HSF. 
Each Krylov subspace generated from an initial fermionic product state $\ket{\psi_0}$ is defined as 
\begin{align}
   &\mathcal K\bigl(H_{\rm PH}, \ket{\psi_0}\bigr) \nonumber\\
   &\ = \mathrm{span}\{
       \ket{\psi_0},
       H_{\rm PH}\ket{\psi_0},
       H_{\rm PH}^{2}\ket{\psi_0},\cdots
   \}. \label{eq:fragment}
\end{align}
Depending on the Krylov subspace, the restricted Hamiltonian can be either integrable or nonintegrable. 

We introduce pseudospin degrees of freedom by grouping each pair of consecutive original sites as 
\begin{align}
    \textrm{Spin} \quad &\ket{\uparrow} \coloneqq \ket{01}, \ \ket{\downarrow} \coloneqq \ket{10}, \\
    \textrm{Fracton} \quad &\ket{+} \coloneqq \ket{11}, \ \ket{-} \coloneqq \ket{00}. 
\end{align}
The corresponding pseudospin-1/2 raising, lowering, and pseudospin-$z$ operators at site $m$ are defined as 
\begin{align}
    \sigma_{m}^+ &= c_{2m-1}^{\dagger}c_{2m}, \\
    \sigma_{m}^- &= c_{2m}^{\dagger}c_{2m-1}, \\
    \sigma_m^{z} &= n_{2m} - n_{2m-1} \label{eq:pseudo_z}. 
\end{align}
In what follows, we assume an even system size $L$ and denote the number of pseudospin sites as $N=L/2$. 

We discuss the integrable sector of the pair-hopping Hamiltonian. 
We define $\Pi$ by a projection to the Krylov subspace [Eq.~\eqref{eq:fragment}] generated by product states $\ket{\psi_0}$ composed solely of $\uparrow$ and $\downarrow$. The projection operator satisfies $[\Pi, H_{\rm PH}] = 0$. 
Within this subspace, the pair-hopping Hamiltonian can be mapped exactly onto the integrable spin-1/2 XX Hamiltonian:
\begin{align}
    \Pi H_{\rm PH}\Pi = H_{XX} = J\sum\limits_{m=1}^{N-1} (\sigma_{m}^{+}\sigma_{m+1}^{-} + h.c.), 
\end{align}
here we call this subspace as the integrable sector. 
Applying the Jordan--Wigner transformation, we further map the XX model onto the tight-binding model for spinless fermions: 
\begin{align}
     H_{XX} = J\sum\limits_{m=1}^{N-1} (d_{m}^{\dagger}d_{m+1} + h.c. ), 
     \label{eq:tight-bindingmodel}
\end{align}
where $d_m^{\dagger} = \exp(i\pi \sum_{l < m} \sigma_l^{+}\sigma_l^{-} ) \sigma_m^{+}$, $d_m = \exp(i\pi \sum_{l < m} \sigma_l^{+}\sigma_l^{-} ) \sigma_m^{-}$ are the creation and annihilation operators of a spinless fermion at site $m$, defined in terms of the pseudospin operators. 
Furthermore, we diagonalize the Hamiltonian in Eq.~\eqref{eq:tight-bindingmodel} using the discrete sine transform~\cite{buschTightbindingElectronsOpen1987}
\begin{align}
    d_{m}^{\dagger} = \sqrt{\frac{2}{N+1}}\sum\limits_{\lambda=1}^{N} \sin\left(\frac{\lambda m \pi}{N+1}\right) d_{\lambda}^{\dagger}, \nonumber\\
    d_{m} = \sqrt{\frac{2}{N+1}}\sum\limits_{\lambda=1}^{N} \sin\left(\frac{\lambda m \pi}{N+1}\right) d_{\lambda}, 
\end{align}
which yields the diagonal form
 \begin{align}
    H_{XX} = \sum\limits_{\lambda=1}^{N} \epsilon_{\lambda} d_{\lambda}^{\dagger} d_{\lambda}. 
    \label{eq:XXmodel_wavenumber}
\end{align}
where the energy spectrum is given by $\epsilon_{\lambda} = 2J\cos \left(\lambda\pi/(N+1)\right)$ for $\lambda = 1, \dots, N$. 
Note that when the initial state contains two or more consecutive identical fractons, such as $++\dots +$ or $--\dots -$ in the middle, these regions can act as domain walls. 
Thus we can disconnect the other regions from one another, depicted as frozen domains in Fig.~\ref{fig:hsfqsp_all}. 

\section{Quantum signal processing meets Hilbert space fragmentation}~\label{sec4} 
In this section, we propose the nonequilibrium control protocol for the 1D pair-hopping model and examine how QSP operates beyond conventional settings by leveraging HSF. Our driving protocol is based on the following time-dependent Hamiltonian:
\begin{align}
    H(t) = &H_{\rm PH}(t) + H_{\rm stag}(t),
    \label{eq:qsp_total}
\end{align}
with
\begin{align}
    H_{\rm PH}(t) = &J(t) \sum\limits_{j=1}^{L-3} (c_{j}^{\dagger}c_{j+3}^{\dagger}c_{j+2}c_{j+1} + h.c.), \nonumber\\
    H_{\rm stag}(t) = &\frac{h(t)}{2} \sum_{m=1}^{N} (-1)^{m} (n_{2m} - n_{2m-1}), 
\end{align}
where the first term corresponds to the 1D pair-hopping Hamiltonian [Eq.~\eqref{eq:pair-hopping}] and the second term represents the staggered on-site potential with a four-fold periodicity. 
The amplitudes $J(t)$ and $h(t)$ depends on time $t$. 
We employ a piecewise-constant driving protocol in which $H_{\rm PH}$ and $H_{\rm stag}$ are switched on alternately. 
We begin with an initial evolution under $H_{\rm stag}$ realized by setting $J(t)=0$ and $h(t) = h$ for a duration $t_0$. We then apply $d$ layers, each consisting of an evolution under $H_{\rm PH}$, with $J(t)=J$ and $h(t) = 0$ for a fixed time $t^{\prime}$, followed by an evolution under $H_{\rm stag}$, $J(t)=0$ and $h(t) = h$ for a duration $t_r$ $(r=1, \dots, d)$. 
The resulting time evolution is given by 
\begin{align}
    &U^{\rm Real}_{h\vec{t}} \nonumber\\
    &= \exp(-it_{0}H_{\rm stag}) \prod_{r=1}^{d} \bigl[ \exp(-i t^{\prime} H_{\rm PH}) \exp(-i t_{r}H_{\rm stag}) \bigr]. 
    \label{eq:bdgsu2qsp_pairhopping_real}
\end{align}
Below, we show that the unitary operator introduced above realizes QSP in the integrable sector. 
We note that $H_{\rm stag}$ respects the same symmetries as $H_{XX}$ described in Eqs.~\eqref{eq:particlenumberconservation} and \eqref{eq:comconservation} and that its action does not disrupt the frozen regions. As a result, the HSF structure is preserved, namely, $[\Pi, H(t)]=0$. In the following, we therefore discuss Eq.~\eqref{eq:bdgsu2qsp_pairhopping_real} separately in the integrable and nonintegrable sectors.

\subsection{QSP in the integrable sector}
We analyze the behavior of $H(t)$ in the integrable sectors and prove that QSP can be performed in these sectors. 
In the previous QSP framework for the 1D TFIM~\cite{bastidasQuantumSignalProcessing2024}, QSP was implemented by exploiting the integrable structure of the system. Concretely, the 1D TFIM can be mapped to noninteracting fermions described by a quadratic Hamiltonian, known as the BdG Hamiltonian, via the Jordan--Wigner transformation. The dynamics then decomposes into BdG subsectors labeled by the momentum, and QSP operates within each subsector.  

In contrast, for generic interacting many-body systems, the number of local conserved quantities does not scale with system size, and a mapping to a quadratic Hamiltonian is not guaranteed. 
Moreover, nonintegrable dynamics are typically quantum chaotic and tend to thermalize, accompanied by the spreading of information and entanglement from local to global degrees of freedom.
To address these challenges and provide a QSP framework for nonintegrable systems, one natural approach is to embed a nonthermal subspace within an HSF system and perform QSP within that subsector. 

We begin by transforming Eq.~\eqref{eq:XXmodel_wavenumber} into BdG form by introducing the spinor representation: 
\begin{align}
\Psi_{\lambda} \coloneqq &\textrm{H}_{\rm ad} 
\begin{pmatrix} 
d_{\lambda} \\ 
d_{N+1-\lambda} 
\end{pmatrix},  
\end{align}
where $\textrm{H}_{\rm ad} = \begin{pmatrix} 1 & 1 \\ 1 & -1 \end{pmatrix}/\sqrt{2}$ denotes the Hadamard gate. 
Assuming for simplicity that $N$ is even, we obtain 
\begin{align}
        H_{XX} = &\sum_{\lambda = 1}^{N/2} \epsilon_{\lambda} \Psi_{\lambda}^{\dagger} X \Psi_{\lambda}. 
            \label{eq:tb-diagonalization}
\end{align}
Thus, the free evolution generated by $H_{XX}$ corresponds to an $X$ rotation in each BdG subspace. 
In contrast, the on-site staggered potential with a four-fold periodicity yields a BdG Hamiltonian proportional to $Z$ and therefore generates a $Z$ rotation:
\begin{align}
    H_{\rm stag} 
= & -\sum_{\lambda = 1}^{N/2} h \Psi_{\lambda}^{\dagger} Z \Psi_{\lambda}. 
     \label{eq:staggered-potential}
\end{align}
Therefore, Eq.~\eqref{eq:qsp_total} in the integrable sector can be written as
\begin{align}
    \Pi H(t) \Pi = & \Pi (H_{\rm PH}(t) + H_{\rm stag}(t))\Pi \nonumber\\
    = &\sum_{\lambda = 1}^{N/2} \epsilon_{\lambda}(t) \Psi_{\lambda}^{\dagger} X \Psi_{\lambda} - \sum_{\lambda = 1}^{N/2} h(t) \Psi_{\lambda}^{\dagger} Z\Psi_{\lambda}, 
\end{align}
where $\epsilon_{\lambda}(t) = 2J(t)\cos \left(\lambda\pi/(N+1)\right)$. 
Switching $J(t)$ and $h(t)$ corresponds to toggling between the configurations $\epsilon_{\lambda}(t)=\epsilon_{\lambda}, h(t) = 0$ and $\epsilon_{\lambda}(t)=0, h(t) = h$. The resulting time evolution in each BdG sector realizes the QSP sequence in Eq.~\eqref{eq:qsp-single}:
\begin{align}
    \Pi U^{\rm Real}_{h\vec{t}} \Pi = &e^{-it_{0}\Pi H_{\rm stag} \Pi} \prod_{r=1}^{d} \bigl[e^{-i t^{\prime} H_{XX} } e^{-i t_r \Pi H_{\rm stag}\Pi} \bigr] \nonumber\\
    = &\bigotimes_{\lambda} U_{h\vec{t}}(a_{\lambda}), 
    \label{eq:bdgsu2qsp_pairhopping_real_integrable}
\end{align}
with 
\begin{align}
    &U_{h\vec{t}} (a_{\lambda}) = e^{iht_{0}Z} \prod_{r=1}^{d} \bigl[ e^{i \cos^{-1}a_{\lambda} X} e^{i ht_{r}Z} \bigr], 
    \label{eq:bdg-su2-qsp}
\end{align}
where $a_{\lambda}$ is the signal satisfying the following condition: 
\begin{equation} 
    \epsilon_{\lambda}t^{\prime} = -\cos^{-1}a_{\lambda}.
\end{equation}
In this study, we impose $Jt^{\prime} = -\pi/2$ to obtain $a_{\lambda} = \cos (\pi \cos \left(\lambda\pi/(N+1)\right))$. Equation~\eqref{eq:bdg-su2-qsp} corresponds to the polynomial matrix analogous to Eq.~\eqref{eq:qsp_poly}: 
\begin{align}
    U_{PQ}(a_{\lambda})  =
    \begin{pmatrix}
        P(a_{\lambda}) & i Q(a_{\lambda}) \sqrt{1 - a_{\lambda}^2} \\
        i Q^*(a_{\lambda}) \sqrt{1 - a_{\lambda}^2} & P^*(a_\lambda)
      \end{pmatrix}
      \label{eq:bdgqsp_poly}, 
\end{align}
where $P(a_{\lambda})$ and $Q(a_{\lambda})$ denote complex polynomials of $a_{\lambda}$ satisfying conditions in Eqs.~\eqref{eq:degree}--\eqref{eq:unitarity}. 
Therefore, by choosing an appropriate phase sequence $h\vec{t}$, it is possible to implement the unitary evolution in Eq.~\eqref{eq:bdgqsp_poly}, which satisfies the required condition in each BdG subspace. 

Our protocol is richer than the previous strategy for the 1D TFIM~\cite{bastidasQuantumSignalProcessing2024}. 
The key point is that the HSF structure allows us to select the relevant sector of the Hilbert space depending on the initial product-state configuration, thereby fully separating integrable and nonintegrable sectors. 
As a concrete initial state, we consider a Néel state interrupted by a single domain wall $++\cdots +$: 
    \begin{align}
        \ket{\psi(0)} = &\ket{\uparrow \downarrow \uparrow \downarrow \cdots \uparrow \downarrow ++\cdots + \uparrow \downarrow \cdots \uparrow \downarrow}. 
    \end{align}
The left and right regions divided by the domain wall are independently emulated via QSP based on Eq.~\eqref{eq:bdg-su2-qsp}.
In this way, by exploiting the HSF structure, we can perform simultaneous, parallel emulation of nonequilibrium dynamics using the staggered potential (Fig.~\ref{fig:hsfqsp_all}).

\subsection{Transition probabilities in the integrable sector}
Here, we show that the polynomial transformation implemented by QSP enables flexible control of the dynamics in the integrable sector. To this end, we evaluate the transition probabilities generated by appropriately tuned pulse sequences. 
We prepare the Néel state as the initial state in the integrable sector, which corresponds to the $\ket{0}_{\lambda}$ state in pseudospin-$z$ basis of each BdG sector: 
\begin{align}
    \ket{\psi(0)} = &\ket{\uparrow \downarrow \uparrow \downarrow  \cdots  \uparrow \downarrow} \nonumber\\
    = &\bigotimes_{\lambda=1}^{N/2} \ket{0}_{\lambda}. 
\end{align}
The transition probability is given by 
\begin{align}  
    |\braket{\psi(0)| \Pi U^{\rm Real}_{h\vec{t}}\Pi | \psi(0)}| ^{2}
    = &\prod_{\lambda=1}^{N/2} |\bra{0}_{\lambda} U_{h\vec{t}}(a_{\lambda})\ket{0}_{\lambda}|^{2} \nonumber\\
    = &\prod_{\lambda=1}^{N/2} |P(a_{\lambda})|^{2}, 
\end{align}
showing that the polynomial transformation acts independently on each BdG sector. 
\begin{figure}
\includegraphics[width=3.4in]{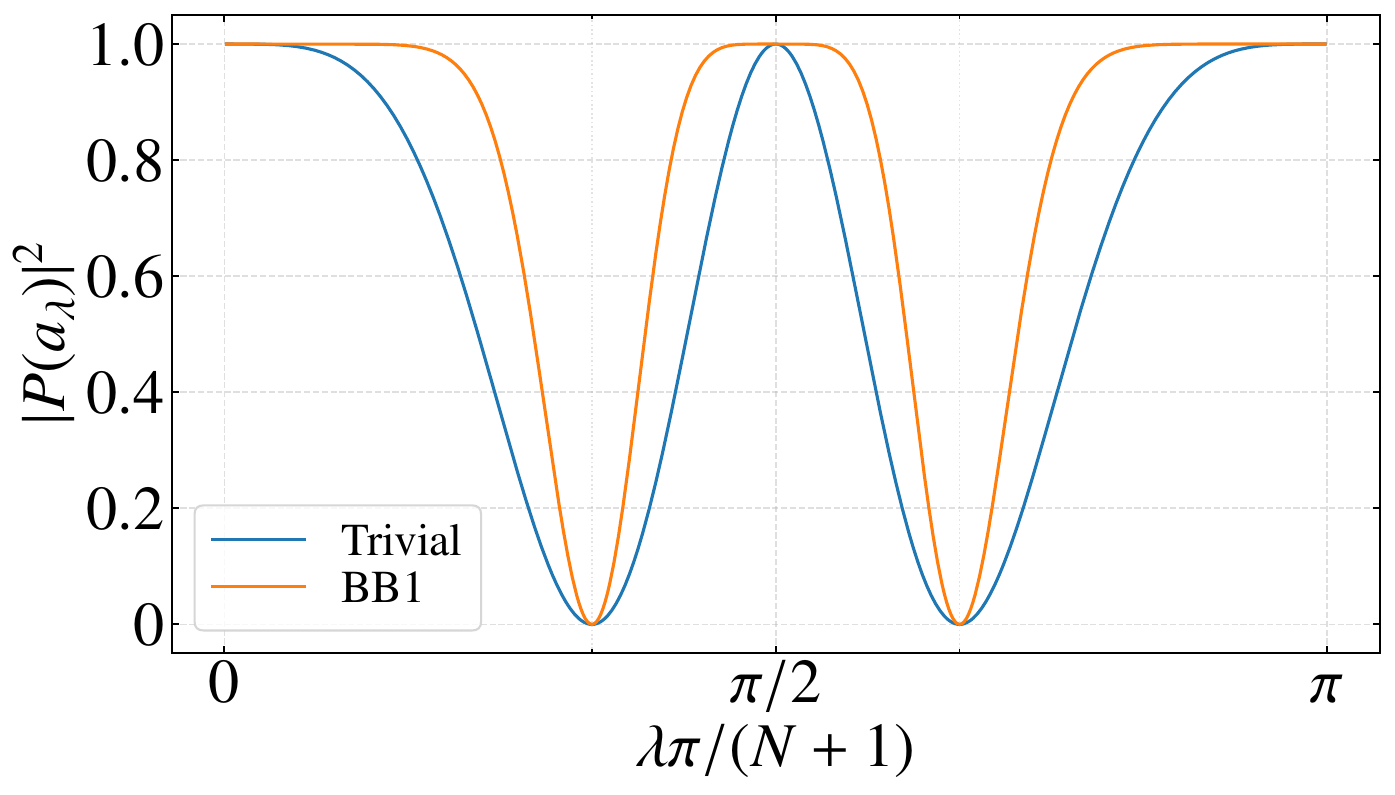}
\caption{Transition probabilities for each BdG sector and their dependence on the phase sequences. The blue curve depicts the transition probability for the trivial sequence $h\vec{t} = (0, 0)$. The orange curve depicts the transition probability for the BB1 pulse sequence $h\vec{t} = (\pi/2, -\chi, 2\chi, 0, -2\chi, \chi)$, where $\chi = \frac{1}{2} \cos^{-1} (-1/4)$.}
\label{fig:result_pulsetest}
\end{figure}
In Fig.~\ref{fig:result_pulsetest}, we depict the 
transition probability for each BdG sector, for two different phase sequences. 
The quantity $|P(a_{\lambda})|^{2}$ is plotted as a function of $\lambda \pi/(N+1)$.
The simplest example is the trivial phase sequence, given by $h\vec{t} = (0, 0)$, which yields the Chebyshev polynomial of the first kind $P(a_{\lambda}) = a_{\lambda}$~\cite{martynEfficientFullycoherentQuantum2023, bastidasQuantumSignalProcessing2024}. 
On the other hand, we consider the BB1 pulse sequence~\cite{martynGrandUnificationQuantum2021, bastidasQuantumSignalProcessing2024, vandersypenNMRTechniquesQuantum2005}, 
\begin{align}
      h\vec{t} = (\pi/2, -\chi, 2\chi, 0, -2\chi, \chi)  \eqqcolon \vec{\phi}_{\rm BB1}, \label{eq:bb1}
\end{align}
as the nontrivial phase sequence, where we set $\chi = \frac{1}{2} \cos^{-1} (-1/4)$.  
As shown in the figure, similar to conventional QSP~\cite{martynGrandUnificationQuantum2021}, the BB1 sequence of phases yields robust peaks at $\lambda \pi/(N+1) = 0, \pi/2, \pi$, as well as sharp feature at $\lambda \pi/(N+1) = \pi/3$ and $2\pi/3$. 
Since the signal $a_\lambda$ depends nonlinearly on $\lambda$ through a cosine function, the resulting peaks appear broader.

\subsection{Demonstrating thermalization}
In this subsection, we examine the real-time dynamics of Eq.~\eqref{eq:bdgsu2qsp_pairhopping_real} to elucidate thermalization in nonintegrable sectors. For the simulations below, we compute the quantum many-body dynamics using the \texttt{QuSpin} Python package~\cite{weinbergQuSpinPythonPackage2017, weinbergQuSpinPythonPackage2019}.

\begin{figure*}
        \includegraphics[width=7in]{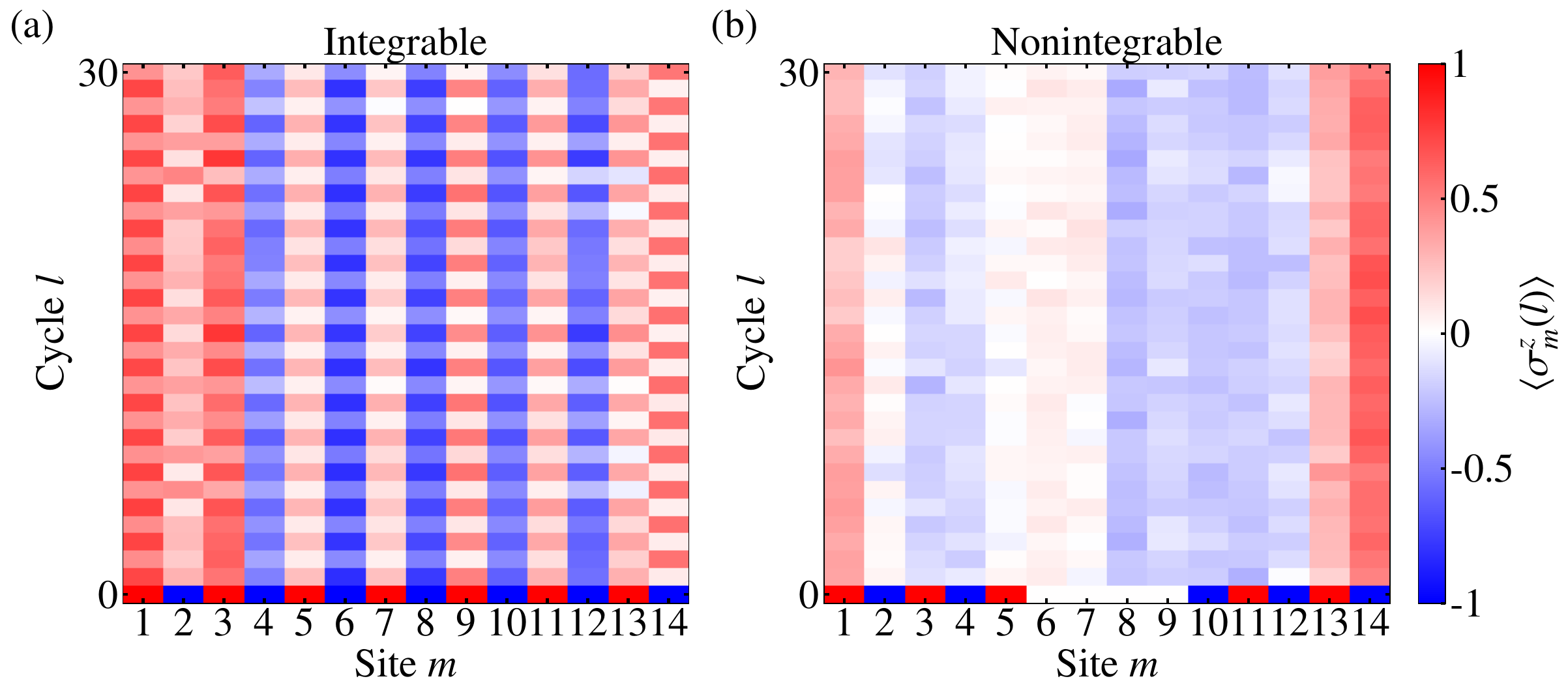}
    \caption{
       Stroboscopic time evolution (a) for an initial state in the integrable sector $\ket{\uparrow \downarrow \uparrow \downarrow \uparrow \downarrow \uparrow \downarrow \uparrow \downarrow \uparrow \downarrow \uparrow \downarrow}$ , and (b) for an initial state in the nonintegrable sector $\ket{\uparrow \downarrow \uparrow \downarrow \uparrow -++-\downarrow \uparrow \downarrow\uparrow \downarrow}$.
    The horizontal axis denotes the site index $m$ in the pseudospin representation, and the vertical axis denotes the cycle number $l$ of the BB1 sequences. The color scale shows the expectation value of the pseudospin-$z$ operator.   
    }
    \label{fig:result_thermalization1}
\end{figure*}
We perform numerical calculations of the time evolution of local observables to demonstrate the coexistence of QSP in the integrable sector and thermalization in the nonintegrable sector within a single system.
As shown by our analytical results above, QSP can be applied to initial states in the integrable sector, enabling the realization of desired polynomial transformations. 
On the other hand, when we prepare initial states in the nonintegrable sector, we cannot map them onto the desired Hamiltonian; thus, the QSP in the BdG subsector is not well defined.
However, the operation described by Eq.~\eqref{eq:bdgsu2qsp_pairhopping_real} is invariant regardless of the choice of initial state. 
Therefore, by applying the same protocol in Eq.~\eqref{eq:bdgsu2qsp_pairhopping_real} to initial states in sectors with different integrability, we can compare the behavior of local observables when QSP is realized in integrable systems with that when the same QSP sequence is applied in nonintegrable systems.

We take the system size to be $L=28$ fermion sites, or equivalently $N=14$ pseudospin sites.
As the initial state in the integrable sector, we use the Néel state $\ket{\uparrow \downarrow \uparrow \downarrow \uparrow \downarrow \uparrow \downarrow \uparrow \downarrow \uparrow \downarrow \uparrow \downarrow}$. 
On the other hand, as the initial state in the nonintegrable sector, we use $\ket{\uparrow \downarrow \uparrow \downarrow \uparrow -++-\downarrow \uparrow \downarrow\uparrow \downarrow}$~\cite{moudgalyaThermalizationItsAbsence2020}.
To discuss thermalization in a manner analogous to that used for periodically driven systems, we repeat the BB1 pulse sequence of Eq.~\eqref{eq:bb1} in each cycle $l$. 
We compute the stroboscopic dynamics of the expectation value of Eq.~\eqref{eq:pseudo_z}:
\begin{align}
    \braket{\sigma_m^z (l)} &= \braket{\psi(0)|\left[
    \bigl(U^{\rm Real}_{\vec{\phi}_{\rm BB1}}\bigr)^\dagger
    \right]^l \sigma_m^z (U^{\rm Real}_{\vec{\phi}_{\rm BB1}})^l|\psi (0)},
\end{align}
and compare the expectation value with the infinite-temperature value.
Within the HSF sector, the infinite-temperature expectation value of the pseudospin-$z$ operator in Eq.~\eqref{eq:pseudo_z} is given by the ensemble average over that sector:
\begin{align}
    \braket{\sigma_{m}^{z}} \coloneqq \frac{1}{D} \sum_{i=1}^{D} \braket{\phi_{i} | \sigma_{m}^{z} | \phi_{i}}, 
    \label{eq:infinitetemp}
\end{align}
where $\{\ket{\phi_{i}}\}_{i=1,\ldots, D}$ is an orthonormal basis spanning the Krylov subspace in Eq.~\eqref{eq:fragment} generated from the initial state by the pair-hopping Hamiltonian, and $D$ is the dimension of the Krylov subspace.

Fig.~\ref{fig:result_thermalization1} shows numerical results for the pseudospin dynamics of initial states in the integrable and nonintegrable sectors.  
For the integrable-sector initial state presented in Fig.~\ref{fig:result_thermalization1}(a), the pseudospin expectation values at the bulk sites show little decay over the calculated timescales and remain significantly different between adjacent sites, indicating that spatial inhomogeneity is preserved. 
In contrast, for the nonintegrable-sector initial state presented in Fig.~\ref{fig:result_thermalization1}(b), the bulk pseudospin expectation values approach spatially homogeneous over the calculated timescales. 
This result suggests Krylov-restricted thermalization to the infinite-temperature value defined in Eq.~\eqref{eq:infinitetemp}~\cite{iadecolaSymmetryFragmentation2025}.
\begin{figure*}
    \begin{center}
        \includegraphics[width=7in]{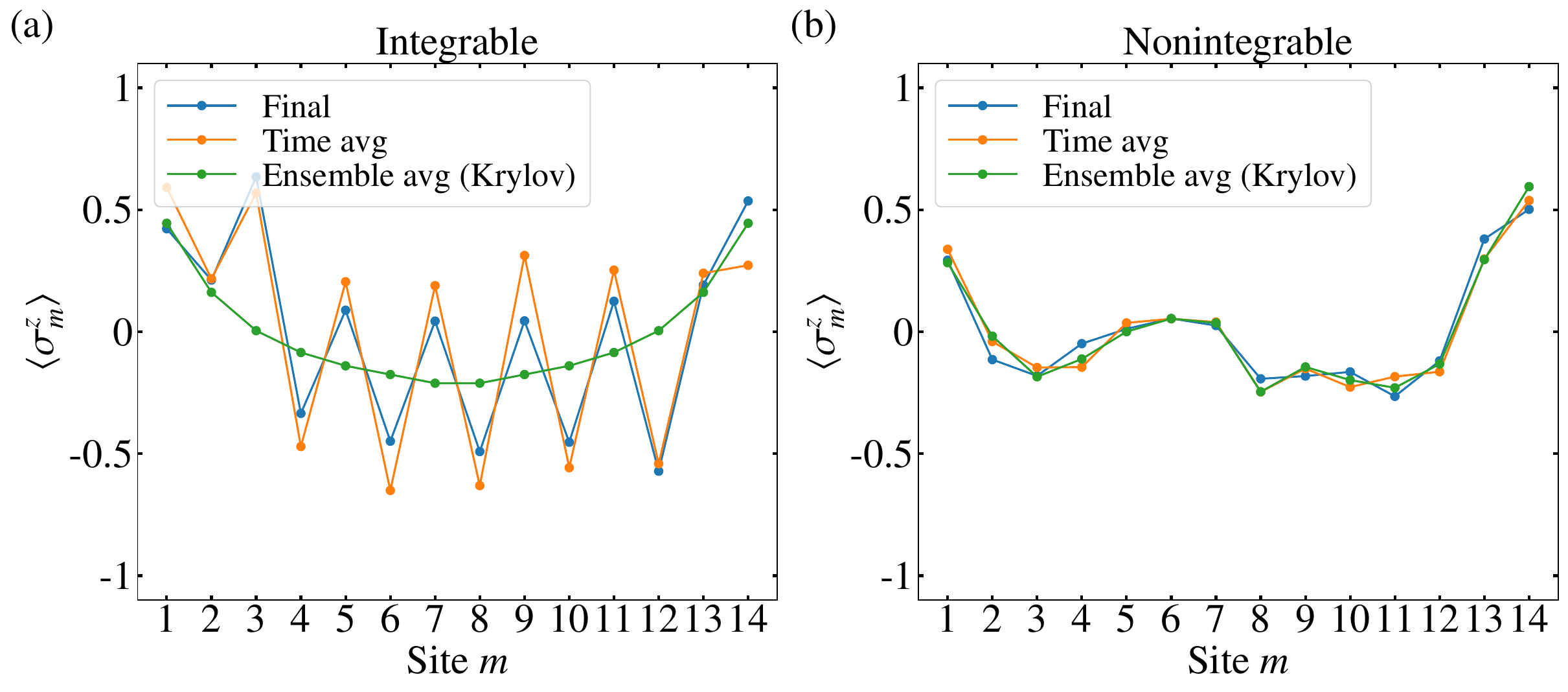}
    \end{center}
    \caption{Comparison of the expectation values of the pseudospin operator $\sigma_{m}^{z}$ at the final time (Final), time-averaged expectation values (Time avg), and ensemble-averaged expectation values over the Krylov subspace (Ensemble avg), for initial states (a) in the integrable sector $\ket{\uparrow \downarrow \uparrow \downarrow \uparrow \downarrow \uparrow \downarrow \uparrow \downarrow \uparrow \downarrow \uparrow \downarrow}$ and (b) the nonintegrable sector $\ket{\uparrow \downarrow \uparrow \downarrow \uparrow -++-\downarrow \uparrow \downarrow\uparrow \downarrow}$ under the repitition of BB1 sequences.
    The horizontal axis denotes the site index $m$ in the pseudospin representation, and the vertical axis denotes the expectation value of the pseudospin-$z$ operator.}
    \label{fig:result_thermalization2}
\end{figure*}
Accordingly, Fig.~\ref{fig:result_thermalization2} shows the expectation values of the pseudospin at the final time (cycle $l = 30$), the time-averaged expectation values (i.e., the diagonal-ensemble values), and the ensemble average over the Krylov subspace generated by applying the pair-hopping model to the initial state. 
We can see from Fig.~\ref{fig:result_thermalization2}(a) that, in the integrable sector, the final-state pseudospin expectation values are close to the time-averaged ones, indicating relaxation of the local observables.
By contrast, the time-averaged values do not converge to the ensemble averages, indicating that thermalization does not occur with respect to the pseudospin expectation values. 
We also note that since the HSF structure is preserved, QSP control remains protected from coupling to other nonintegrable sectors. 
By contrast, in the nonintegrable sector shown in Fig.~\ref{fig:result_thermalization2}(b), both the final-state pseudospin expectation values and the time-averaged expectation values show nearly identical values across all the bulk sites. 
This indicates that, in the nonintegrable sector, the dynamics are consistent with Krylov-space-restricted thermalization over the simulated timescales. 
These results reveal two distinct behaviors based on the choice of initial state configuration: a regime that does not thermalize under QSP control and a regime showing signatures of thermalization.
In general, even in the system that exhibits thermalization, the ability to execute QSP solely through the choice of initial state goes beyond prior work on integrable systems~\cite{bastidasQuantumSignalProcessing2024}.


\section{Conclusion}~\label{sec5}
In summary, we have investigated the implementation of QSP in the integrable sector of a one-dimensional pair-hopping model that exhibits HSF.
By alternating the time evolutions generated by the pair-hopping Hamiltonian and a four-fold staggered potential, we can realize QSP in each BdG subspace. 
Our scheme reveals that the time evolution generated by a broad class of BdG-integrable Hamiltonians can be realized through quantum emulation within an integrable sector embedded in an HSF system.
As in prior work by Bastidas \textit{et al.}~\cite{bastidasQuantumSignalProcessing2024}, our framework does not require any additional nonlocal operations or ancillary resources. 
Moreover, by introducing domain walls, the system can be separated into independent regions, thereby enabling parallel emulation across multiple disconnected integrable regions. 
As a result, implementing QSP in a fragmented system suggests a route toward scalable QSP beyond conventional frameworks based solely on integrable sectors.
Moreover, the HSF structure allows integrable and nonintegrable sectors to coexist within a single system. 
Under the same QSP driving sequence, the initial state in the integrable sector does not exhibit thermalization in the pseudospin expectation values, whereas the initial state in the nonintegrable sector suggests Krylov-space-restricted thermalization.
These results show that HSF is not merely a constraint on dynamics, but can also provide a useful tool for implementing and organizing QSP in many-body systems. 

An important direction for future work is the application of QSP to higher-dimensional systems. 
Extending QSP-based emulation to higher dimensions would substantially broaden the range of experimentally relevant phenomena that become programmable, including frustrated magnetism and topological phases that are intrinsically two or three-dimensional. 
It is also of interest to ask whether implementations of QSP can be extended beyond BdG integrability to target more complex Bethe-integrable systems or even nonintegrable systems. 
Finally, our results motivate further generalization of QSP in many-body systems beyond the standard setting~\cite{motlaghGeneralizedQuantumSignal2024}. 
The recent work has generalized QSP by allowing arbitrary SU(2) signal-processing rotations, suggesting that the class of achievable polynomial transformations could be substantially expanded. 



\section{Acknowledgement}
The authors thank A. Ono and D. Sasamoto for fruitful discussions. 
Parts of the numerical calculations were performed in the supercomputing systems in ISSP, the University of Tokyo.
K. M. is supported by JST PRESTO Grant No. JPMJPR235A, JSPS KAKENHI Grant No. JP24K16974, and JST Moonshot R\&D Grant No. JPMJMS2061. J. N. is supported by JSPS KAKENHI Grant No. JP23H04865, JP24K00563. 




\bibliography{references(2)}

\begin{thebibliography}{65}%
\makeatletter
\providecommand \@ifxundefined [1]{%
 \@ifx{#1\undefined}
}%
\providecommand \@ifnum [1]{%
 \ifnum #1\expandafter \@firstoftwo
 \else \expandafter \@secondoftwo
 \fi
}%
\providecommand \@ifx [1]{%
 \ifx #1\expandafter \@firstoftwo
 \else \expandafter \@secondoftwo
 \fi
}%
\providecommand \natexlab [1]{#1}%
\providecommand \enquote  [1]{``#1''}%
\providecommand \bibnamefont  [1]{#1}%
\providecommand \bibfnamefont [1]{#1}%
\providecommand \citenamefont [1]{#1}%
\providecommand \href@noop [0]{\@secondoftwo}%
\providecommand \href [0]{\begingroup \@sanitize@url \@href}%
\providecommand \@href[1]{\@@startlink{#1}\@@href}%
\providecommand \@@href[1]{\endgroup#1\@@endlink}%
\providecommand \@sanitize@url [0]{\catcode `\\12\catcode `\$12\catcode `\&12\catcode `\#12\catcode `\^12\catcode `\_12\catcode `\%12\relax}%
\providecommand \@@startlink[1]{}%
\providecommand \@@endlink[0]{}%
\providecommand \url  [0]{\begingroup\@sanitize@url \@url }%
\providecommand \@url [1]{\endgroup\@href {#1}{\urlprefix }}%
\providecommand \urlprefix  [0]{URL }%
\providecommand \Eprint [0]{\href }%
\providecommand \doibase [0]{https://doi.org/}%
\providecommand \selectlanguage [0]{\@gobble}%
\providecommand \bibinfo  [0]{\@secondoftwo}%
\providecommand \bibfield  [0]{\@secondoftwo}%
\providecommand \translation [1]{[#1]}%
\providecommand \BibitemOpen [0]{}%
\providecommand \bibitemStop [0]{}%
\providecommand \bibitemNoStop [0]{.\EOS\space}%
\providecommand \EOS [0]{\spacefactor3000\relax}%
\providecommand \BibitemShut  [1]{\csname bibitem#1\endcsname}%
\let\auto@bib@innerbib\@empty
\bibitem [{\citenamefont {Deutsch}(1991)}]{deutschQuantumStatisticalMechanics1991}%
  \BibitemOpen
  \bibfield  {author} {\bibinfo {author} {\bibfnamefont {J.~M.}\ \bibnamefont {Deutsch}},\ }\href {https://doi.org/10.1103/PhysRevA.43.2046} {\bibfield  {journal} {\bibinfo  {journal} {Phys. Rev. A}\ }\textbf {\bibinfo {volume} {43}},\ \bibinfo {pages} {2046} (\bibinfo {year} {1991})}\BibitemShut {NoStop}%
\bibitem [{\citenamefont {Srednicki}(1994)}]{srednickiChaosQuantumThermalization1994}%
  \BibitemOpen
  \bibfield  {author} {\bibinfo {author} {\bibfnamefont {M.}~\bibnamefont {Srednicki}},\ }\href {https://doi.org/10.1103/PhysRevE.50.888} {\bibfield  {journal} {\bibinfo  {journal} {Phys. Rev. E}\ }\textbf {\bibinfo {volume} {50}},\ \bibinfo {pages} {888} (\bibinfo {year} {1994})}\BibitemShut {NoStop}%
\bibitem [{\citenamefont {Rigol}\ \emph {et~al.}(2008)\citenamefont {Rigol}, \citenamefont {Dunjko},\ and\ \citenamefont {Olshanii}}]{rigolThermalizationItsMechanism2008}%
  \BibitemOpen
  \bibfield  {author} {\bibinfo {author} {\bibfnamefont {M.}~\bibnamefont {Rigol}}, \bibinfo {author} {\bibfnamefont {V.}~\bibnamefont {Dunjko}},\ and\ \bibinfo {author} {\bibfnamefont {M.}~\bibnamefont {Olshanii}},\ }\href {https://doi.org/10.1038/nature06838} {\bibfield  {journal} {\bibinfo  {journal} {Nature}\ }\textbf {\bibinfo {volume} {452}},\ \bibinfo {pages} {854} (\bibinfo {year} {2008})}\BibitemShut {NoStop}%
\bibitem [{\citenamefont {Rigol}(2009)}]{rigolBreakdownThermalizationFinite2009}%
  \BibitemOpen
  \bibfield  {author} {\bibinfo {author} {\bibfnamefont {M.}~\bibnamefont {Rigol}},\ }\href {https://doi.org/10.1103/PhysRevLett.103.100403} {\bibfield  {journal} {\bibinfo  {journal} {Phys. Rev. Lett.}\ }\textbf {\bibinfo {volume} {103}},\ \bibinfo {pages} {100403} (\bibinfo {year} {2009})}\BibitemShut {NoStop}%
\bibitem [{\citenamefont {D'Alessio}\ \emph {et~al.}(2016)\citenamefont {D'Alessio}, \citenamefont {Kafri}, \citenamefont {Polkovnikov},\ and\ \citenamefont {Rigol}}]{dalessioQuantumChaosEigenstate2016}%
  \BibitemOpen
  \bibfield  {author} {\bibinfo {author} {\bibfnamefont {L.}~\bibnamefont {D'Alessio}}, \bibinfo {author} {\bibfnamefont {Y.}~\bibnamefont {Kafri}}, \bibinfo {author} {\bibfnamefont {A.}~\bibnamefont {Polkovnikov}},\ and\ \bibinfo {author} {\bibfnamefont {M.}~\bibnamefont {Rigol}},\ }\href {https://doi.org/10.1080/00018732.2016.1198134} {\bibfield  {journal} {\bibinfo  {journal} {Advances in Physics}\ }\textbf {\bibinfo {volume} {65}},\ \bibinfo {pages} {239} (\bibinfo {year} {2016})}\BibitemShut {NoStop}%
\bibitem [{\citenamefont {Mori}\ \emph {et~al.}(2018)\citenamefont {Mori}, \citenamefont {Ikeda}, \citenamefont {Kaminishi},\ and\ \citenamefont {Ueda}}]{moriThermalizationPrethermalizationIsolated2018}%
  \BibitemOpen
  \bibfield  {author} {\bibinfo {author} {\bibfnamefont {T.}~\bibnamefont {Mori}}, \bibinfo {author} {\bibfnamefont {T.~N.}\ \bibnamefont {Ikeda}}, \bibinfo {author} {\bibfnamefont {E.}~\bibnamefont {Kaminishi}},\ and\ \bibinfo {author} {\bibfnamefont {M.}~\bibnamefont {Ueda}},\ }\href {https://doi.org/10.1088/1361-6455/aabcdf} {\bibfield  {journal} {\bibinfo  {journal} {J. Phys. B: At. Mol. Opt. Phys.}\ }\textbf {\bibinfo {volume} {51}},\ \bibinfo {pages} {112001} (\bibinfo {year} {2018})}\BibitemShut {NoStop}%
\bibitem [{\citenamefont {Sugimoto}\ \emph {et~al.}(2025)\citenamefont {Sugimoto}, \citenamefont {Hamazaki},\ and\ \citenamefont {Ueda}}]{sugimotoBoundsEigenstateThermalization2025}%
  \BibitemOpen
  \bibfield  {author} {\bibinfo {author} {\bibfnamefont {S.}~\bibnamefont {Sugimoto}}, \bibinfo {author} {\bibfnamefont {R.}~\bibnamefont {Hamazaki}},\ and\ \bibinfo {author} {\bibfnamefont {M.}~\bibnamefont {Ueda}},\ }\href {https://doi.org/10.48550/arXiv.2303.10069} {\bibinfo {title} {Bounds on eigenstate thermalization}} (\bibinfo {year} {2025}),\ \Eprint {https://arxiv.org/abs/2303.10069} {arXiv:2303.10069 [cond-mat]} \BibitemShut {NoStop}%
\bibitem [{\citenamefont {Oganesyan}\ and\ \citenamefont {Huse}(2007)}]{oganesyanLocalizationInteractingFermions2007}%
  \BibitemOpen
  \bibfield  {author} {\bibinfo {author} {\bibfnamefont {V.}~\bibnamefont {Oganesyan}}\ and\ \bibinfo {author} {\bibfnamefont {D.~A.}\ \bibnamefont {Huse}},\ }\href {https://doi.org/10.1103/PhysRevB.75.155111} {\bibfield  {journal} {\bibinfo  {journal} {Phys. Rev. B}\ }\textbf {\bibinfo {volume} {75}},\ \bibinfo {pages} {155111} (\bibinfo {year} {2007})}\BibitemShut {NoStop}%
\bibitem [{\citenamefont {Pal}\ and\ \citenamefont {Huse}(2010)}]{palManybodyLocalizationPhase2010}%
  \BibitemOpen
  \bibfield  {author} {\bibinfo {author} {\bibfnamefont {A.}~\bibnamefont {Pal}}\ and\ \bibinfo {author} {\bibfnamefont {D.~A.}\ \bibnamefont {Huse}},\ }\href {https://doi.org/10.1103/PhysRevB.82.174411} {\bibfield  {journal} {\bibinfo  {journal} {Phys. Rev. B}\ }\textbf {\bibinfo {volume} {82}},\ \bibinfo {pages} {174411} (\bibinfo {year} {2010})}\BibitemShut {NoStop}%
\bibitem [{\citenamefont {Serbyn}\ \emph {et~al.}(2013)\citenamefont {Serbyn}, \citenamefont {Papi{\'c}},\ and\ \citenamefont {Abanin}}]{serbynLocalConservationLaws2013a}%
  \BibitemOpen
  \bibfield  {author} {\bibinfo {author} {\bibfnamefont {M.}~\bibnamefont {Serbyn}}, \bibinfo {author} {\bibfnamefont {Z.}~\bibnamefont {Papi{\'c}}},\ and\ \bibinfo {author} {\bibfnamefont {D.~A.}\ \bibnamefont {Abanin}},\ }\href {https://doi.org/10.1103/PhysRevLett.111.127201} {\bibfield  {journal} {\bibinfo  {journal} {Phys. Rev. Lett.}\ }\textbf {\bibinfo {volume} {111}},\ \bibinfo {pages} {127201} (\bibinfo {year} {2013})}\BibitemShut {NoStop}%
\bibitem [{\citenamefont {Huse}\ \emph {et~al.}(2014)\citenamefont {Huse}, \citenamefont {Nandkishore},\ and\ \citenamefont {Oganesyan}}]{husePhenomenologyFullyManybodylocalized2014}%
  \BibitemOpen
  \bibfield  {author} {\bibinfo {author} {\bibfnamefont {D.~A.}\ \bibnamefont {Huse}}, \bibinfo {author} {\bibfnamefont {R.}~\bibnamefont {Nandkishore}},\ and\ \bibinfo {author} {\bibfnamefont {V.}~\bibnamefont {Oganesyan}},\ }\href {https://doi.org/10.1103/PhysRevB.90.174202} {\bibfield  {journal} {\bibinfo  {journal} {Phys. Rev. B}\ }\textbf {\bibinfo {volume} {90}},\ \bibinfo {pages} {174202} (\bibinfo {year} {2014})}\BibitemShut {NoStop}%
\bibitem [{\citenamefont {Luitz}\ \emph {et~al.}(2015)\citenamefont {Luitz}, \citenamefont {Laflorencie},\ and\ \citenamefont {Alet}}]{luitzManybodyLocalizationEdge2015}%
  \BibitemOpen
  \bibfield  {author} {\bibinfo {author} {\bibfnamefont {D.~J.}\ \bibnamefont {Luitz}}, \bibinfo {author} {\bibfnamefont {N.}~\bibnamefont {Laflorencie}},\ and\ \bibinfo {author} {\bibfnamefont {F.}~\bibnamefont {Alet}},\ }\href {https://doi.org/10.1103/PhysRevB.91.081103} {\bibfield  {journal} {\bibinfo  {journal} {Phys. Rev. B}\ }\textbf {\bibinfo {volume} {91}},\ \bibinfo {pages} {081103} (\bibinfo {year} {2015})}\BibitemShut {NoStop}%
\bibitem [{\citenamefont {Nandkishore}\ and\ \citenamefont {Huse}(2015)}]{nandkishoreManyBodyLocalizationThermalization2015}%
  \BibitemOpen
  \bibfield  {author} {\bibinfo {author} {\bibfnamefont {R.}~\bibnamefont {Nandkishore}}\ and\ \bibinfo {author} {\bibfnamefont {D.~A.}\ \bibnamefont {Huse}},\ }\href {https://doi.org/10.1146/annurev-conmatphys-031214-014726} {\bibfield  {journal} {\bibinfo  {journal} {Annu. Rev. Condens. Matter Phys.}\ }\textbf {\bibinfo {volume} {6}},\ \bibinfo {pages} {15} (\bibinfo {year} {2015})}\BibitemShut {NoStop}%
\bibitem [{\citenamefont {Abanin}\ \emph {et~al.}(2019)\citenamefont {Abanin}, \citenamefont {Altman}, \citenamefont {Bloch},\ and\ \citenamefont {Serbyn}}]{abaninColloquiumManybodyLocalization2019}%
  \BibitemOpen
  \bibfield  {author} {\bibinfo {author} {\bibfnamefont {D.~A.}\ \bibnamefont {Abanin}}, \bibinfo {author} {\bibfnamefont {E.}~\bibnamefont {Altman}}, \bibinfo {author} {\bibfnamefont {I.}~\bibnamefont {Bloch}},\ and\ \bibinfo {author} {\bibfnamefont {M.}~\bibnamefont {Serbyn}},\ }\href {https://doi.org/10.1103/RevModPhys.91.021001} {\bibfield  {journal} {\bibinfo  {journal} {Rev. Mod. Phys.}\ }\textbf {\bibinfo {volume} {91}},\ \bibinfo {pages} {021001} (\bibinfo {year} {2019})}\BibitemShut {NoStop}%
\bibitem [{\citenamefont {Sierant}\ \emph {et~al.}(2024)\citenamefont {Sierant}, \citenamefont {Lewenstein}, \citenamefont {Scardicchio}, \citenamefont {Vidmar},\ and\ \citenamefont {Zakrzewski}}]{sierantManyBodyLocalizationAge2024}%
  \BibitemOpen
  \bibfield  {author} {\bibinfo {author} {\bibfnamefont {P.}~\bibnamefont {Sierant}}, \bibinfo {author} {\bibfnamefont {M.}~\bibnamefont {Lewenstein}}, \bibinfo {author} {\bibfnamefont {A.}~\bibnamefont {Scardicchio}}, \bibinfo {author} {\bibfnamefont {L.}~\bibnamefont {Vidmar}},\ and\ \bibinfo {author} {\bibfnamefont {J.}~\bibnamefont {Zakrzewski}},\ }\href {https://doi.org/10.48550/arXiv.2403.07111} {\bibinfo {title} {Many-{{Body Localization}} in the {{Age}} of {{Classical Computing}}}} (\bibinfo {year} {2024}),\ \Eprint {https://arxiv.org/abs/2403.07111} {arXiv:2403.07111} \BibitemShut {NoStop}%
\bibitem [{\citenamefont {Kollar}\ \emph {et~al.}(2011)\citenamefont {Kollar}, \citenamefont {Wolf},\ and\ \citenamefont {Eckstein}}]{kollarGeneralizedGibbsEnsemble2011}%
  \BibitemOpen
  \bibfield  {author} {\bibinfo {author} {\bibfnamefont {M.}~\bibnamefont {Kollar}}, \bibinfo {author} {\bibfnamefont {F.~A.}\ \bibnamefont {Wolf}},\ and\ \bibinfo {author} {\bibfnamefont {M.}~\bibnamefont {Eckstein}},\ }\href {https://doi.org/10.1103/PhysRevB.84.054304} {\bibfield  {journal} {\bibinfo  {journal} {Phys. Rev. B}\ }\textbf {\bibinfo {volume} {84}},\ \bibinfo {pages} {054304} (\bibinfo {year} {2011})}\BibitemShut {NoStop}%
\bibitem [{\citenamefont {Bernien}\ \emph {et~al.}(2017)\citenamefont {Bernien}, \citenamefont {Schwartz}, \citenamefont {Keesling}, \citenamefont {Levine}, \citenamefont {Omran}, \citenamefont {Pichler}, \citenamefont {Choi}, \citenamefont {Zibrov}, \citenamefont {Endres}, \citenamefont {Greiner}, \citenamefont {Vuleti{\'c}},\ and\ \citenamefont {Lukin}}]{bernienProbingManybodyDynamics2017}%
  \BibitemOpen
  \bibfield  {author} {\bibinfo {author} {\bibfnamefont {H.}~\bibnamefont {Bernien}}, \bibinfo {author} {\bibfnamefont {S.}~\bibnamefont {Schwartz}}, \bibinfo {author} {\bibfnamefont {A.}~\bibnamefont {Keesling}}, \bibinfo {author} {\bibfnamefont {H.}~\bibnamefont {Levine}}, \bibinfo {author} {\bibfnamefont {A.}~\bibnamefont {Omran}}, \bibinfo {author} {\bibfnamefont {H.}~\bibnamefont {Pichler}}, \bibinfo {author} {\bibfnamefont {S.}~\bibnamefont {Choi}}, \bibinfo {author} {\bibfnamefont {A.~S.}\ \bibnamefont {Zibrov}}, \bibinfo {author} {\bibfnamefont {M.}~\bibnamefont {Endres}}, \bibinfo {author} {\bibfnamefont {M.}~\bibnamefont {Greiner}}, \bibinfo {author} {\bibfnamefont {V.}~\bibnamefont {Vuleti{\'c}}},\ and\ \bibinfo {author} {\bibfnamefont {M.~D.}\ \bibnamefont {Lukin}},\ }\href {https://doi.org/10.1038/nature24622} {\bibfield  {journal} {\bibinfo  {journal} {Nature}\ }\textbf {\bibinfo {volume} {551}},\ \bibinfo {pages} {579} (\bibinfo {year} {2017})}\BibitemShut {NoStop}%
\bibitem [{\citenamefont {Shiraishi}\ and\ \citenamefont {Mori}(2017)}]{shiraishiSystematicConstructionCounterexamples2017}%
  \BibitemOpen
  \bibfield  {author} {\bibinfo {author} {\bibfnamefont {N.}~\bibnamefont {Shiraishi}}\ and\ \bibinfo {author} {\bibfnamefont {T.}~\bibnamefont {Mori}},\ }\href {https://doi.org/10.1103/PhysRevLett.119.030601} {\bibfield  {journal} {\bibinfo  {journal} {Phys. Rev. Lett.}\ }\textbf {\bibinfo {volume} {119}},\ \bibinfo {pages} {030601} (\bibinfo {year} {2017})}\BibitemShut {NoStop}%
\bibitem [{\citenamefont {Turner}\ \emph {et~al.}(2018)\citenamefont {Turner}, \citenamefont {Michailidis}, \citenamefont {Abanin}, \citenamefont {Serbyn},\ and\ \citenamefont {Papi{\'c}}}]{turnerWeakErgodicityBreaking2018}%
  \BibitemOpen
  \bibfield  {author} {\bibinfo {author} {\bibfnamefont {C.~J.}\ \bibnamefont {Turner}}, \bibinfo {author} {\bibfnamefont {A.~A.}\ \bibnamefont {Michailidis}}, \bibinfo {author} {\bibfnamefont {D.~A.}\ \bibnamefont {Abanin}}, \bibinfo {author} {\bibfnamefont {M.}~\bibnamefont {Serbyn}},\ and\ \bibinfo {author} {\bibfnamefont {Z.}~\bibnamefont {Papi{\'c}}},\ }\href {https://doi.org/10.1038/s41567-018-0137-5} {\bibfield  {journal} {\bibinfo  {journal} {Nat. Phys.}\ }\textbf {\bibinfo {volume} {14}},\ \bibinfo {pages} {745} (\bibinfo {year} {2018})}\BibitemShut {NoStop}%
\bibitem [{\citenamefont {Moudgalya}\ \emph {et~al.}(2020{\natexlab{a}})\citenamefont {Moudgalya}, \citenamefont {Prem}, \citenamefont {Nandkishore}, \citenamefont {Regnault},\ and\ \citenamefont {Bernevig}}]{moudgalyaThermalizationItsAbsence2020}%
  \BibitemOpen
  \bibfield  {author} {\bibinfo {author} {\bibfnamefont {S.}~\bibnamefont {Moudgalya}}, \bibinfo {author} {\bibfnamefont {A.}~\bibnamefont {Prem}}, \bibinfo {author} {\bibfnamefont {R.}~\bibnamefont {Nandkishore}}, \bibinfo {author} {\bibfnamefont {N.}~\bibnamefont {Regnault}},\ and\ \bibinfo {author} {\bibfnamefont {B.~A.}\ \bibnamefont {Bernevig}},\ }in\ \href {https://doi.org/10.1142/9789811231711_0009} {\emph {\bibinfo {booktitle} {Memorial {{Volume}} for {{Shoucheng Zhang}}}}}\ (\bibinfo  {publisher} {WORLD SCIENTIFIC},\ \bibinfo {year} {2020})\ pp.\ \bibinfo {pages} {147--209}\BibitemShut {NoStop}%
\bibitem [{\citenamefont {Mark}\ \emph {et~al.}(2020)\citenamefont {Mark}, \citenamefont {Lin},\ and\ \citenamefont {Motrunich}}]{markUnifiedStructureExact2020}%
  \BibitemOpen
  \bibfield  {author} {\bibinfo {author} {\bibfnamefont {D.~K.}\ \bibnamefont {Mark}}, \bibinfo {author} {\bibfnamefont {C.-J.}\ \bibnamefont {Lin}},\ and\ \bibinfo {author} {\bibfnamefont {O.~I.}\ \bibnamefont {Motrunich}},\ }\href {https://doi.org/10.1103/PhysRevB.101.195131} {\bibfield  {journal} {\bibinfo  {journal} {Phys. Rev. B}\ }\textbf {\bibinfo {volume} {101}},\ \bibinfo {pages} {195131} (\bibinfo {year} {2020})}\BibitemShut {NoStop}%
\bibitem [{\citenamefont {Serbyn}\ \emph {et~al.}(2021)\citenamefont {Serbyn}, \citenamefont {Abanin},\ and\ \citenamefont {Papi{\'c}}}]{serbynQuantumManybodyScars2021}%
  \BibitemOpen
  \bibfield  {author} {\bibinfo {author} {\bibfnamefont {M.}~\bibnamefont {Serbyn}}, \bibinfo {author} {\bibfnamefont {D.~A.}\ \bibnamefont {Abanin}},\ and\ \bibinfo {author} {\bibfnamefont {Z.}~\bibnamefont {Papi{\'c}}},\ }\href {https://doi.org/10.1038/s41567-021-01230-2} {\bibfield  {journal} {\bibinfo  {journal} {Nat. Phys.}\ }\textbf {\bibinfo {volume} {17}},\ \bibinfo {pages} {675} (\bibinfo {year} {2021})}\BibitemShut {NoStop}%
\bibitem [{\citenamefont {Papi{\'c}}(2021)}]{papicWeakErgodicityBreaking2021}%
  \BibitemOpen
  \bibfield  {author} {\bibinfo {author} {\bibfnamefont {Z.}~\bibnamefont {Papi{\'c}}},\ }\href@noop {} {\bibinfo {title} {Weak ergodicity breaking through the lens of quantum entanglement}} (\bibinfo {year} {2021}),\ \Eprint {https://arxiv.org/abs/2108.03460} {arXiv:2108.03460 [cond-mat]} \BibitemShut {NoStop}%
\bibitem [{\citenamefont {Moudgalya}\ \emph {et~al.}(2022{\natexlab{a}})\citenamefont {Moudgalya}, \citenamefont {Bernevig},\ and\ \citenamefont {Regnault}}]{moudgalyaQuantumManyBodyScars2022}%
  \BibitemOpen
  \bibfield  {author} {\bibinfo {author} {\bibfnamefont {S.}~\bibnamefont {Moudgalya}}, \bibinfo {author} {\bibfnamefont {B.~A.}\ \bibnamefont {Bernevig}},\ and\ \bibinfo {author} {\bibfnamefont {N.}~\bibnamefont {Regnault}},\ }\href {https://doi.org/10.1088/1361-6633/ac73a0} {\bibfield  {journal} {\bibinfo  {journal} {Rep. Prog. Phys.}\ }\textbf {\bibinfo {volume} {85}},\ \bibinfo {pages} {086501} (\bibinfo {year} {2022}{\natexlab{a}})}\BibitemShut {NoStop}%
\bibitem [{\citenamefont {Chandran}\ \emph {et~al.}(2023)\citenamefont {Chandran}, \citenamefont {Iadecola}, \citenamefont {Khemani},\ and\ \citenamefont {Moessner}}]{chandranQuantumManyBodyScars2023}%
  \BibitemOpen
  \bibfield  {author} {\bibinfo {author} {\bibfnamefont {A.}~\bibnamefont {Chandran}}, \bibinfo {author} {\bibfnamefont {T.}~\bibnamefont {Iadecola}}, \bibinfo {author} {\bibfnamefont {V.}~\bibnamefont {Khemani}},\ and\ \bibinfo {author} {\bibfnamefont {R.}~\bibnamefont {Moessner}},\ }\href {https://doi.org/10.1146/annurev-conmatphys-031620-101617} {\bibfield  {journal} {\bibinfo  {journal} {Annu. Rev. Condens. Matter Phys.}\ }\textbf {\bibinfo {volume} {14}},\ \bibinfo {pages} {443} (\bibinfo {year} {2023})}\BibitemShut {NoStop}%
\bibitem [{\citenamefont {Else}\ \emph {et~al.}(2016)\citenamefont {Else}, \citenamefont {Bauer},\ and\ \citenamefont {Nayak}}]{elseFloquetTimeCrystals2016}%
  \BibitemOpen
  \bibfield  {author} {\bibinfo {author} {\bibfnamefont {D.~V.}\ \bibnamefont {Else}}, \bibinfo {author} {\bibfnamefont {B.}~\bibnamefont {Bauer}},\ and\ \bibinfo {author} {\bibfnamefont {C.}~\bibnamefont {Nayak}},\ }\href {https://doi.org/10.1103/PhysRevLett.117.090402} {\bibfield  {journal} {\bibinfo  {journal} {Phys. Rev. Lett.}\ }\textbf {\bibinfo {volume} {117}},\ \bibinfo {pages} {090402} (\bibinfo {year} {2016})}\BibitemShut {NoStop}%
\bibitem [{\citenamefont {Yao}\ \emph {et~al.}(2017)\citenamefont {Yao}, \citenamefont {Potter}, \citenamefont {Potirniche},\ and\ \citenamefont {Vishwanath}}]{yaoDiscreteTimeCrystals2017}%
  \BibitemOpen
  \bibfield  {author} {\bibinfo {author} {\bibfnamefont {N.~Y.}\ \bibnamefont {Yao}}, \bibinfo {author} {\bibfnamefont {A.~C.}\ \bibnamefont {Potter}}, \bibinfo {author} {\bibfnamefont {I.-D.}\ \bibnamefont {Potirniche}},\ and\ \bibinfo {author} {\bibfnamefont {A.}~\bibnamefont {Vishwanath}},\ }\href {https://doi.org/10.1103/PhysRevLett.118.030401} {\bibfield  {journal} {\bibinfo  {journal} {Phys. Rev. Lett.}\ }\textbf {\bibinfo {volume} {118}},\ \bibinfo {pages} {030401} (\bibinfo {year} {2017})}\BibitemShut {NoStop}%
\bibitem [{\citenamefont {Khemani}\ \emph {et~al.}(2016)\citenamefont {Khemani}, \citenamefont {Lazarides}, \citenamefont {Moessner},\ and\ \citenamefont {Sondhi}}]{khemaniPhaseStructureDriven2016}%
  \BibitemOpen
  \bibfield  {author} {\bibinfo {author} {\bibfnamefont {V.}~\bibnamefont {Khemani}}, \bibinfo {author} {\bibfnamefont {A.}~\bibnamefont {Lazarides}}, \bibinfo {author} {\bibfnamefont {R.}~\bibnamefont {Moessner}},\ and\ \bibinfo {author} {\bibfnamefont {S.~L.}\ \bibnamefont {Sondhi}},\ }\href {https://doi.org/10.1103/PhysRevLett.116.250401} {\bibfield  {journal} {\bibinfo  {journal} {Phys. Rev. Lett.}\ }\textbf {\bibinfo {volume} {116}},\ \bibinfo {pages} {250401} (\bibinfo {year} {2016})}\BibitemShut {NoStop}%
\bibitem [{\citenamefont {Ippoliti}\ \emph {et~al.}(2021)\citenamefont {Ippoliti}, \citenamefont {Kechedzhi}, \citenamefont {Moessner}, \citenamefont {Sondhi},\ and\ \citenamefont {Khemani}}]{ippolitiManyBodyPhysicsNISQ2021}%
  \BibitemOpen
  \bibfield  {author} {\bibinfo {author} {\bibfnamefont {M.}~\bibnamefont {Ippoliti}}, \bibinfo {author} {\bibfnamefont {K.}~\bibnamefont {Kechedzhi}}, \bibinfo {author} {\bibfnamefont {R.}~\bibnamefont {Moessner}}, \bibinfo {author} {\bibfnamefont {S.}~\bibnamefont {Sondhi}},\ and\ \bibinfo {author} {\bibfnamefont {V.}~\bibnamefont {Khemani}},\ }\href {https://doi.org/10.1103/PRXQuantum.2.030346} {\bibfield  {journal} {\bibinfo  {journal} {PRX Quantum}\ }\textbf {\bibinfo {volume} {2}},\ \bibinfo {pages} {030346} (\bibinfo {year} {2021})}\BibitemShut {NoStop}%
\bibitem [{\citenamefont {Randall}\ \emph {et~al.}(2021)\citenamefont {Randall}, \citenamefont {Bradley}, \citenamefont {{van der Gronden}}, \citenamefont {Galicia}, \citenamefont {Abobeih}, \citenamefont {Markham}, \citenamefont {Twitchen}, \citenamefont {Machado}, \citenamefont {Yao},\ and\ \citenamefont {Taminiau}}]{randallManybodylocalizedDiscrete2021}%
  \BibitemOpen
  \bibfield  {author} {\bibinfo {author} {\bibfnamefont {J.}~\bibnamefont {Randall}}, \bibinfo {author} {\bibfnamefont {C.~E.}\ \bibnamefont {Bradley}}, \bibinfo {author} {\bibfnamefont {F.~V.}\ \bibnamefont {{van der Gronden}}}, \bibinfo {author} {\bibfnamefont {A.}~\bibnamefont {Galicia}}, \bibinfo {author} {\bibfnamefont {M.~H.}\ \bibnamefont {Abobeih}}, \bibinfo {author} {\bibfnamefont {M.}~\bibnamefont {Markham}}, \bibinfo {author} {\bibfnamefont {D.~J.}\ \bibnamefont {Twitchen}}, \bibinfo {author} {\bibfnamefont {F.}~\bibnamefont {Machado}}, \bibinfo {author} {\bibfnamefont {N.~Y.}\ \bibnamefont {Yao}},\ and\ \bibinfo {author} {\bibfnamefont {T.~H.}\ \bibnamefont {Taminiau}},\ }\href {https://doi.org/10.1126/science.abk0603} {\bibfield  {journal} {\bibinfo  {journal} {Science}\ }\textbf {\bibinfo {volume} {374}},\ \bibinfo {pages} {1474} (\bibinfo {year} {2021})}\BibitemShut {NoStop}%
\bibitem [{\citenamefont {Mi}\ \emph {et~al.}(2022)\citenamefont {Mi}, \citenamefont {Ippoliti}, \citenamefont {Quintana}, \citenamefont {Greene}, \citenamefont {Chen}, \citenamefont {Gross}, \citenamefont {Arute}, \citenamefont {Arya}, \citenamefont {Atalaya}, \citenamefont {Babbush}, \citenamefont {Bardin}, \citenamefont {Basso}, \citenamefont {Bengtsson}, \citenamefont {Bilmes}, \citenamefont {Bourassa}, \citenamefont {Brill}, \citenamefont {Broughton}, \citenamefont {Buckley}, \citenamefont {Buell}, \citenamefont {Burkett}, \citenamefont {Bushnell}, \citenamefont {Chiaro}, \citenamefont {Collins}, \citenamefont {Courtney}, \citenamefont {Debroy}, \citenamefont {Demura}, \citenamefont {Derk}, \citenamefont {Dunsworth}, \citenamefont {Eppens}, \citenamefont {Erickson}, \citenamefont {Farhi}, \citenamefont {Fowler}, \citenamefont {Foxen}, \citenamefont {Gidney}, \citenamefont {Giustina}, \citenamefont {Harrigan}, \citenamefont {Harrington}, \citenamefont {Hilton}, \citenamefont {Ho}, \citenamefont {Hong},
  \citenamefont {Huang}, \citenamefont {Huff}, \citenamefont {Huggins}, \citenamefont {Ioffe}, \citenamefont {Isakov}, \citenamefont {Iveland}, \citenamefont {Jeffrey}, \citenamefont {Jiang}, \citenamefont {Jones}, \citenamefont {Kafri}, \citenamefont {Khattar}, \citenamefont {Kim}, \citenamefont {Kitaev}, \citenamefont {Klimov}, \citenamefont {Korotkov}, \citenamefont {Kostritsa}, \citenamefont {Landhuis}, \citenamefont {Laptev}, \citenamefont {Lee}, \citenamefont {Lee}, \citenamefont {Locharla}, \citenamefont {Lucero}, \citenamefont {Martin}, \citenamefont {McClean}, \citenamefont {McCourt}, \citenamefont {McEwen}, \citenamefont {Miao}, \citenamefont {Mohseni}, \citenamefont {Montazeri}, \citenamefont {Mruczkiewicz}, \citenamefont {Naaman}, \citenamefont {Neeley}, \citenamefont {Neill}, \citenamefont {Newman}, \citenamefont {Niu}, \citenamefont {O'Brien}, \citenamefont {Opremcak}, \citenamefont {Ostby}, \citenamefont {Pato}, \citenamefont {Petukhov}, \citenamefont {Rubin}, \citenamefont {Sank},
  \citenamefont {Satzinger}, \citenamefont {Shvarts}, \citenamefont {Su}, \citenamefont {Strain}, \citenamefont {Szalay}, \citenamefont {Trevithick}, \citenamefont {Villalonga}, \citenamefont {White}, \citenamefont {Yao}, \citenamefont {Yeh}, \citenamefont {Yoo}, \citenamefont {Zalcman}, \citenamefont {Neven}, \citenamefont {Boixo}, \citenamefont {Smelyanskiy}, \citenamefont {Megrant}, \citenamefont {Kelly}, \citenamefont {Chen}, \citenamefont {Sondhi}, \citenamefont {Moessner}, \citenamefont {Kechedzhi}, \citenamefont {Khemani},\ and\ \citenamefont {Roushan}}]{miTimecrystallineEigenstateOrder2022}%
  \BibitemOpen
  \bibfield  {author} {\bibinfo {author} {\bibfnamefont {X.}~\bibnamefont {Mi}}, \bibinfo {author} {\bibfnamefont {M.}~\bibnamefont {Ippoliti}}, \bibinfo {author} {\bibfnamefont {C.}~\bibnamefont {Quintana}}, \bibinfo {author} {\bibfnamefont {A.}~\bibnamefont {Greene}}, \bibinfo {author} {\bibfnamefont {Z.}~\bibnamefont {Chen}}, \bibinfo {author} {\bibfnamefont {J.}~\bibnamefont {Gross}}, \bibinfo {author} {\bibfnamefont {F.}~\bibnamefont {Arute}}, \bibinfo {author} {\bibfnamefont {K.}~\bibnamefont {Arya}}, \bibinfo {author} {\bibfnamefont {J.}~\bibnamefont {Atalaya}}, \bibinfo {author} {\bibfnamefont {R.}~\bibnamefont {Babbush}}, \bibinfo {author} {\bibfnamefont {J.~C.}\ \bibnamefont {Bardin}}, \bibinfo {author} {\bibfnamefont {J.}~\bibnamefont {Basso}}, \bibinfo {author} {\bibfnamefont {A.}~\bibnamefont {Bengtsson}}, \bibinfo {author} {\bibfnamefont {A.}~\bibnamefont {Bilmes}}, \bibinfo {author} {\bibfnamefont {A.}~\bibnamefont {Bourassa}}, \bibinfo {author} {\bibfnamefont {L.}~\bibnamefont {Brill}},
  \bibinfo {author} {\bibfnamefont {M.}~\bibnamefont {Broughton}}, \bibinfo {author} {\bibfnamefont {B.~B.}\ \bibnamefont {Buckley}}, \bibinfo {author} {\bibfnamefont {D.~A.}\ \bibnamefont {Buell}}, \bibinfo {author} {\bibfnamefont {B.}~\bibnamefont {Burkett}}, \bibinfo {author} {\bibfnamefont {N.}~\bibnamefont {Bushnell}}, \bibinfo {author} {\bibfnamefont {B.}~\bibnamefont {Chiaro}}, \bibinfo {author} {\bibfnamefont {R.}~\bibnamefont {Collins}}, \bibinfo {author} {\bibfnamefont {W.}~\bibnamefont {Courtney}}, \bibinfo {author} {\bibfnamefont {D.}~\bibnamefont {Debroy}}, \bibinfo {author} {\bibfnamefont {S.}~\bibnamefont {Demura}}, \bibinfo {author} {\bibfnamefont {A.~R.}\ \bibnamefont {Derk}}, \bibinfo {author} {\bibfnamefont {A.}~\bibnamefont {Dunsworth}}, \bibinfo {author} {\bibfnamefont {D.}~\bibnamefont {Eppens}}, \bibinfo {author} {\bibfnamefont {C.}~\bibnamefont {Erickson}}, \bibinfo {author} {\bibfnamefont {E.}~\bibnamefont {Farhi}}, \bibinfo {author} {\bibfnamefont {A.~G.}\ \bibnamefont {Fowler}},
  \bibinfo {author} {\bibfnamefont {B.}~\bibnamefont {Foxen}}, \bibinfo {author} {\bibfnamefont {C.}~\bibnamefont {Gidney}}, \bibinfo {author} {\bibfnamefont {M.}~\bibnamefont {Giustina}}, \bibinfo {author} {\bibfnamefont {M.~P.}\ \bibnamefont {Harrigan}}, \bibinfo {author} {\bibfnamefont {S.~D.}\ \bibnamefont {Harrington}}, \bibinfo {author} {\bibfnamefont {J.}~\bibnamefont {Hilton}}, \bibinfo {author} {\bibfnamefont {A.}~\bibnamefont {Ho}}, \bibinfo {author} {\bibfnamefont {S.}~\bibnamefont {Hong}}, \bibinfo {author} {\bibfnamefont {T.}~\bibnamefont {Huang}}, \bibinfo {author} {\bibfnamefont {A.}~\bibnamefont {Huff}}, \bibinfo {author} {\bibfnamefont {W.~J.}\ \bibnamefont {Huggins}}, \bibinfo {author} {\bibfnamefont {L.~B.}\ \bibnamefont {Ioffe}}, \bibinfo {author} {\bibfnamefont {S.~V.}\ \bibnamefont {Isakov}}, \bibinfo {author} {\bibfnamefont {J.}~\bibnamefont {Iveland}}, \bibinfo {author} {\bibfnamefont {E.}~\bibnamefont {Jeffrey}}, \bibinfo {author} {\bibfnamefont {Z.}~\bibnamefont {Jiang}}, \bibinfo
  {author} {\bibfnamefont {C.}~\bibnamefont {Jones}}, \bibinfo {author} {\bibfnamefont {D.}~\bibnamefont {Kafri}}, \bibinfo {author} {\bibfnamefont {T.}~\bibnamefont {Khattar}}, \bibinfo {author} {\bibfnamefont {S.}~\bibnamefont {Kim}}, \bibinfo {author} {\bibfnamefont {A.}~\bibnamefont {Kitaev}}, \bibinfo {author} {\bibfnamefont {P.~V.}\ \bibnamefont {Klimov}}, \bibinfo {author} {\bibfnamefont {A.~N.}\ \bibnamefont {Korotkov}}, \bibinfo {author} {\bibfnamefont {F.}~\bibnamefont {Kostritsa}}, \bibinfo {author} {\bibfnamefont {D.}~\bibnamefont {Landhuis}}, \bibinfo {author} {\bibfnamefont {P.}~\bibnamefont {Laptev}}, \bibinfo {author} {\bibfnamefont {J.}~\bibnamefont {Lee}}, \bibinfo {author} {\bibfnamefont {K.}~\bibnamefont {Lee}}, \bibinfo {author} {\bibfnamefont {A.}~\bibnamefont {Locharla}}, \bibinfo {author} {\bibfnamefont {E.}~\bibnamefont {Lucero}}, \bibinfo {author} {\bibfnamefont {O.}~\bibnamefont {Martin}}, \bibinfo {author} {\bibfnamefont {J.~R.}\ \bibnamefont {McClean}}, \bibinfo {author}
  {\bibfnamefont {T.}~\bibnamefont {McCourt}}, \bibinfo {author} {\bibfnamefont {M.}~\bibnamefont {McEwen}}, \bibinfo {author} {\bibfnamefont {K.~C.}\ \bibnamefont {Miao}}, \bibinfo {author} {\bibfnamefont {M.}~\bibnamefont {Mohseni}}, \bibinfo {author} {\bibfnamefont {S.}~\bibnamefont {Montazeri}}, \bibinfo {author} {\bibfnamefont {W.}~\bibnamefont {Mruczkiewicz}}, \bibinfo {author} {\bibfnamefont {O.}~\bibnamefont {Naaman}}, \bibinfo {author} {\bibfnamefont {M.}~\bibnamefont {Neeley}}, \bibinfo {author} {\bibfnamefont {C.}~\bibnamefont {Neill}}, \bibinfo {author} {\bibfnamefont {M.}~\bibnamefont {Newman}}, \bibinfo {author} {\bibfnamefont {M.~Y.}\ \bibnamefont {Niu}}, \bibinfo {author} {\bibfnamefont {T.~E.}\ \bibnamefont {O'Brien}}, \bibinfo {author} {\bibfnamefont {A.}~\bibnamefont {Opremcak}}, \bibinfo {author} {\bibfnamefont {E.}~\bibnamefont {Ostby}}, \bibinfo {author} {\bibfnamefont {B.}~\bibnamefont {Pato}}, \bibinfo {author} {\bibfnamefont {A.}~\bibnamefont {Petukhov}}, \bibinfo {author}
  {\bibfnamefont {N.~C.}\ \bibnamefont {Rubin}}, \bibinfo {author} {\bibfnamefont {D.}~\bibnamefont {Sank}}, \bibinfo {author} {\bibfnamefont {K.~J.}\ \bibnamefont {Satzinger}}, \bibinfo {author} {\bibfnamefont {V.}~\bibnamefont {Shvarts}}, \bibinfo {author} {\bibfnamefont {Y.}~\bibnamefont {Su}}, \bibinfo {author} {\bibfnamefont {D.}~\bibnamefont {Strain}}, \bibinfo {author} {\bibfnamefont {M.}~\bibnamefont {Szalay}}, \bibinfo {author} {\bibfnamefont {M.~D.}\ \bibnamefont {Trevithick}}, \bibinfo {author} {\bibfnamefont {B.}~\bibnamefont {Villalonga}}, \bibinfo {author} {\bibfnamefont {T.}~\bibnamefont {White}}, \bibinfo {author} {\bibfnamefont {Z.~J.}\ \bibnamefont {Yao}}, \bibinfo {author} {\bibfnamefont {P.}~\bibnamefont {Yeh}}, \bibinfo {author} {\bibfnamefont {J.}~\bibnamefont {Yoo}}, \bibinfo {author} {\bibfnamefont {A.}~\bibnamefont {Zalcman}}, \bibinfo {author} {\bibfnamefont {H.}~\bibnamefont {Neven}}, \bibinfo {author} {\bibfnamefont {S.}~\bibnamefont {Boixo}}, \bibinfo {author} {\bibfnamefont
  {V.}~\bibnamefont {Smelyanskiy}}, \bibinfo {author} {\bibfnamefont {A.}~\bibnamefont {Megrant}}, \bibinfo {author} {\bibfnamefont {J.}~\bibnamefont {Kelly}}, \bibinfo {author} {\bibfnamefont {Y.}~\bibnamefont {Chen}}, \bibinfo {author} {\bibfnamefont {S.~L.}\ \bibnamefont {Sondhi}}, \bibinfo {author} {\bibfnamefont {R.}~\bibnamefont {Moessner}}, \bibinfo {author} {\bibfnamefont {K.}~\bibnamefont {Kechedzhi}}, \bibinfo {author} {\bibfnamefont {V.}~\bibnamefont {Khemani}},\ and\ \bibinfo {author} {\bibfnamefont {P.}~\bibnamefont {Roushan}},\ }\href {https://doi.org/10.1038/s41586-021-04257-w} {\bibfield  {journal} {\bibinfo  {journal} {Nature}\ }\textbf {\bibinfo {volume} {601}},\ \bibinfo {pages} {531} (\bibinfo {year} {2022})}\BibitemShut {NoStop}%
\bibitem [{\citenamefont {Abanin}\ \emph {et~al.}(2017)\citenamefont {Abanin}, \citenamefont {De~Roeck}, \citenamefont {Ho},\ and\ \citenamefont {Huveneers}}]{abaninRigorousTheoryManyBody2017}%
  \BibitemOpen
  \bibfield  {author} {\bibinfo {author} {\bibfnamefont {D.}~\bibnamefont {Abanin}}, \bibinfo {author} {\bibfnamefont {W.}~\bibnamefont {De~Roeck}}, \bibinfo {author} {\bibfnamefont {W.~W.}\ \bibnamefont {Ho}},\ and\ \bibinfo {author} {\bibfnamefont {F.}~\bibnamefont {Huveneers}},\ }\href {https://doi.org/10.1007/s00220-017-2930-x} {\bibfield  {journal} {\bibinfo  {journal} {Commun. Math. Phys.}\ }\textbf {\bibinfo {volume} {354}},\ \bibinfo {pages} {809} (\bibinfo {year} {2017})}\BibitemShut {NoStop}%
\bibitem [{\citenamefont {Mori}\ \emph {et~al.}(2021)\citenamefont {Mori}, \citenamefont {Zhao}, \citenamefont {Mintert}, \citenamefont {Knolle},\ and\ \citenamefont {Moessner}}]{moriRigorousBoundsHeating2021}%
  \BibitemOpen
  \bibfield  {author} {\bibinfo {author} {\bibfnamefont {T.}~\bibnamefont {Mori}}, \bibinfo {author} {\bibfnamefont {H.}~\bibnamefont {Zhao}}, \bibinfo {author} {\bibfnamefont {F.}~\bibnamefont {Mintert}}, \bibinfo {author} {\bibfnamefont {J.}~\bibnamefont {Knolle}},\ and\ \bibinfo {author} {\bibfnamefont {R.}~\bibnamefont {Moessner}},\ }\href {https://doi.org/10.1103/PhysRevLett.127.050602} {\bibfield  {journal} {\bibinfo  {journal} {Phys. Rev. Lett.}\ }\textbf {\bibinfo {volume} {127}},\ \bibinfo {pages} {050602} (\bibinfo {year} {2021})}\BibitemShut {NoStop}%
\bibitem [{\citenamefont {Kuwahara}\ \emph {et~al.}(2016)\citenamefont {Kuwahara}, \citenamefont {Mori},\ and\ \citenamefont {Saito}}]{kuwaharaFloquetMagnusTheory2016}%
  \BibitemOpen
  \bibfield  {author} {\bibinfo {author} {\bibfnamefont {T.}~\bibnamefont {Kuwahara}}, \bibinfo {author} {\bibfnamefont {T.}~\bibnamefont {Mori}},\ and\ \bibinfo {author} {\bibfnamefont {K.}~\bibnamefont {Saito}},\ }\href {https://doi.org/10.1016/j.aop.2016.01.012} {\bibfield  {journal} {\bibinfo  {journal} {Ann. Phys.}\ }\textbf {\bibinfo {volume} {367}},\ \bibinfo {pages} {96} (\bibinfo {year} {2016})}\BibitemShut {NoStop}%
\bibitem [{\citenamefont {Marvian}\ and\ \citenamefont {Lloyd}(2025)}]{marvianEfficientQuantumEmulation2025}%
  \BibitemOpen
  \bibfield  {author} {\bibinfo {author} {\bibfnamefont {I.}~\bibnamefont {Marvian}}\ and\ \bibinfo {author} {\bibfnamefont {S.}~\bibnamefont {Lloyd}},\ }\href {https://doi.org/10.1103/PRXQuantum.6.030346} {\bibfield  {journal} {\bibinfo  {journal} {PRX Quantum}\ }\textbf {\bibinfo {volume} {6}},\ \bibinfo {pages} {030346} (\bibinfo {year} {2025})}\BibitemShut {NoStop}%
\bibitem [{\citenamefont {Feynman}(1982)}]{feynmanSimulatingPhysicsComputers1982}%
  \BibitemOpen
  \bibfield  {author} {\bibinfo {author} {\bibfnamefont {R.~P.}\ \bibnamefont {Feynman}},\ }\href {https://doi.org/10.1007/BF02650179} {\bibfield  {journal} {\bibinfo  {journal} {Int. J. Theor. Phys.}\ }\textbf {\bibinfo {volume} {21}},\ \bibinfo {pages} {467} (\bibinfo {year} {1982})}\BibitemShut {NoStop}%
\bibitem [{\citenamefont {Lloyd}(1996)}]{lloydUniversalQuantumSimulators1996}%
  \BibitemOpen
  \bibfield  {author} {\bibinfo {author} {\bibfnamefont {S.}~\bibnamefont {Lloyd}},\ }\href {https://doi.org/10.1126/science.273.5278.1073} {\bibfield  {journal} {\bibinfo  {journal} {Science}\ }\textbf {\bibinfo {volume} {273}},\ \bibinfo {pages} {1073} (\bibinfo {year} {1996})}\BibitemShut {NoStop}%
\bibitem [{\citenamefont {Georgescu}(2014)}]{georgescuQuantumSimulation2014}%
  \BibitemOpen
  \bibfield  {author} {\bibinfo {author} {\bibfnamefont {I.~M.}\ \bibnamefont {Georgescu}},\ }\href {https://doi.org/10.1103/RevModPhys.86.153} {\bibfield  {journal} {\bibinfo  {journal} {Rev. Mod. Phys.}\ }\textbf {\bibinfo {volume} {86}},\ \bibinfo {pages} {153} (\bibinfo {year} {2014})}\BibitemShut {NoStop}%
\bibitem [{\citenamefont {Bastidas}\ \emph {et~al.}(2024)\citenamefont {Bastidas}, \citenamefont {Zeytino{\u g}lu}, \citenamefont {Rossi}, \citenamefont {Chuang},\ and\ \citenamefont {Munro}}]{bastidasQuantumSignalProcessing2024}%
  \BibitemOpen
  \bibfield  {author} {\bibinfo {author} {\bibfnamefont {V.~M.}\ \bibnamefont {Bastidas}}, \bibinfo {author} {\bibfnamefont {S.}~\bibnamefont {Zeytino{\u g}lu}}, \bibinfo {author} {\bibfnamefont {Z.~M.}\ \bibnamefont {Rossi}}, \bibinfo {author} {\bibfnamefont {I.~L.}\ \bibnamefont {Chuang}},\ and\ \bibinfo {author} {\bibfnamefont {W.~J.}\ \bibnamefont {Munro}},\ }\href {https://doi.org/10.1103/PhysRevB.109.014306} {\bibfield  {journal} {\bibinfo  {journal} {Phys. Rev. B}\ }\textbf {\bibinfo {volume} {109}},\ \bibinfo {pages} {014306} (\bibinfo {year} {2024})}\BibitemShut {NoStop}%
\bibitem [{\citenamefont {Low}\ \emph {et~al.}(2016)\citenamefont {Low}, \citenamefont {Yoder},\ and\ \citenamefont {Chuang}}]{lowMethodologyResonantEquiangular2016}%
  \BibitemOpen
  \bibfield  {author} {\bibinfo {author} {\bibfnamefont {G.~H.}\ \bibnamefont {Low}}, \bibinfo {author} {\bibfnamefont {T.~J.}\ \bibnamefont {Yoder}},\ and\ \bibinfo {author} {\bibfnamefont {I.~L.}\ \bibnamefont {Chuang}},\ }\href {https://doi.org/10.1103/PhysRevX.6.041067} {\bibfield  {journal} {\bibinfo  {journal} {Phys. Rev. X}\ }\textbf {\bibinfo {volume} {6}},\ \bibinfo {pages} {041067} (\bibinfo {year} {2016})}\BibitemShut {NoStop}%
\bibitem [{\citenamefont {Martyn}\ \emph {et~al.}(2021)\citenamefont {Martyn}, \citenamefont {Rossi}, \citenamefont {Tan},\ and\ \citenamefont {Chuang}}]{martynGrandUnificationQuantum2021}%
  \BibitemOpen
  \bibfield  {author} {\bibinfo {author} {\bibfnamefont {J.~M.}\ \bibnamefont {Martyn}}, \bibinfo {author} {\bibfnamefont {Z.~M.}\ \bibnamefont {Rossi}}, \bibinfo {author} {\bibfnamefont {A.~K.}\ \bibnamefont {Tan}},\ and\ \bibinfo {author} {\bibfnamefont {I.~L.}\ \bibnamefont {Chuang}},\ }\href {https://doi.org/10.1103/PRXQuantum.2.040203} {\bibfield  {journal} {\bibinfo  {journal} {PRX Quantum}\ }\textbf {\bibinfo {volume} {2}},\ \bibinfo {pages} {040203} (\bibinfo {year} {2021})}\BibitemShut {NoStop}%
\bibitem [{\citenamefont {Gily{\'e}n}\ \emph {et~al.}(2019)\citenamefont {Gily{\'e}n}, \citenamefont {Su}, \citenamefont {Low},\ and\ \citenamefont {Wiebe}}]{gilyenQuantumSingularValue2019}%
  \BibitemOpen
  \bibfield  {author} {\bibinfo {author} {\bibfnamefont {A.}~\bibnamefont {Gily{\'e}n}}, \bibinfo {author} {\bibfnamefont {Y.}~\bibnamefont {Su}}, \bibinfo {author} {\bibfnamefont {G.~H.}\ \bibnamefont {Low}},\ and\ \bibinfo {author} {\bibfnamefont {N.}~\bibnamefont {Wiebe}},\ }in\ \href {https://doi.org/10.1145/3313276.3316366} {\emph {\bibinfo {booktitle} {Proceedings of the 51st {{Annual ACM SIGACT Symposium}} on {{Theory}} of {{Computing}}}}},\ \bibinfo {series and number} {{{STOC}} 2019}\ (\bibinfo  {publisher} {Association for Computing Machinery},\ \bibinfo {address} {New York, NY, USA},\ \bibinfo {year} {2019})\ pp.\ \bibinfo {pages} {193--204}\BibitemShut {NoStop}%
\bibitem [{\citenamefont {Grover}(1996)}]{groverFastQuantumMechanical1996}%
  \BibitemOpen
  \bibfield  {author} {\bibinfo {author} {\bibfnamefont {L.~K.}\ \bibnamefont {Grover}},\ }in\ \href {https://doi.org/10.1145/237814.237866} {\emph {\bibinfo {booktitle} {Proceedings of the Twenty-Eighth Annual {{ACM}} Symposium on {{Theory}} of {{Computing}}}}},\ \bibinfo {series and number} {{{STOC}} '96}\ (\bibinfo  {publisher} {Association for Computing Machinery},\ \bibinfo {address} {New York, NY, USA},\ \bibinfo {year} {1996})\ pp.\ \bibinfo {pages} {212--219}\BibitemShut {NoStop}%
\bibitem [{\citenamefont {Brassard}\ \emph {et~al.}(2000)\citenamefont {Brassard}, \citenamefont {Hoyer}, \citenamefont {Mosca},\ and\ \citenamefont {Tapp}}]{brassardQuantumAmplitudeAmplification2000}%
  \BibitemOpen
  \bibfield  {author} {\bibinfo {author} {\bibfnamefont {G.}~\bibnamefont {Brassard}}, \bibinfo {author} {\bibfnamefont {P.}~\bibnamefont {Hoyer}}, \bibinfo {author} {\bibfnamefont {M.}~\bibnamefont {Mosca}},\ and\ \bibinfo {author} {\bibfnamefont {A.}~\bibnamefont {Tapp}},\ }\bibfield  {journal} {\bibinfo  {journal} {AMS Contemporary Mathematics Series}\ }\textbf {\bibinfo {volume} {305}},\ \href {https://doi.org/10.1090/conm/305/05215} {10.1090/conm/305/05215} (\bibinfo {year} {2000})\BibitemShut {NoStop}%
\bibitem [{\citenamefont {Rall}\ and\ \citenamefont {Fuller}(2023)}]{rallAmplitudeEstimationQuantum2023}%
  \BibitemOpen
  \bibfield  {author} {\bibinfo {author} {\bibfnamefont {P.}~\bibnamefont {Rall}}\ and\ \bibinfo {author} {\bibfnamefont {B.}~\bibnamefont {Fuller}},\ }\href {https://doi.org/10.22331/q-2023-03-02-937} {\bibfield  {journal} {\bibinfo  {journal} {Quantum}\ }\textbf {\bibinfo {volume} {7}},\ \bibinfo {pages} {937} (\bibinfo {year} {2023})},\ \Eprint {https://arxiv.org/abs/2207.08628} {arXiv:2207.08628 [quant-ph]} \BibitemShut {NoStop}%
\bibitem [{\citenamefont {Low}\ and\ \citenamefont {Chuang}(2017)}]{lowOptimalHamiltonianSimulation2017}%
  \BibitemOpen
  \bibfield  {author} {\bibinfo {author} {\bibfnamefont {G.~H.}\ \bibnamefont {Low}}\ and\ \bibinfo {author} {\bibfnamefont {I.~L.}\ \bibnamefont {Chuang}},\ }\href {https://doi.org/10.1103/PhysRevLett.118.010501} {\bibfield  {journal} {\bibinfo  {journal} {Phys. Rev. Lett.}\ }\textbf {\bibinfo {volume} {118}},\ \bibinfo {pages} {010501} (\bibinfo {year} {2017})}\BibitemShut {NoStop}%
\bibitem [{\citenamefont {Low}\ and\ \citenamefont {Chuang}(2019)}]{lowHamiltonianSimulationQubitization2019}%
  \BibitemOpen
  \bibfield  {author} {\bibinfo {author} {\bibfnamefont {G.~H.}\ \bibnamefont {Low}}\ and\ \bibinfo {author} {\bibfnamefont {I.~L.}\ \bibnamefont {Chuang}},\ }\href {https://doi.org/10.22331/q-2019-07-12-163} {\bibfield  {journal} {\bibinfo  {journal} {Quantum}\ }\textbf {\bibinfo {volume} {3}},\ \bibinfo {pages} {163} (\bibinfo {year} {2019})}\BibitemShut {NoStop}%
\bibitem [{\citenamefont {Sala}\ \emph {et~al.}(2020)\citenamefont {Sala}, \citenamefont {Rakovszky}, \citenamefont {Verresen}, \citenamefont {Knap},\ and\ \citenamefont {Pollmann}}]{salaErgodicityBreakingArising2020}%
  \BibitemOpen
  \bibfield  {author} {\bibinfo {author} {\bibfnamefont {P.}~\bibnamefont {Sala}}, \bibinfo {author} {\bibfnamefont {T.}~\bibnamefont {Rakovszky}}, \bibinfo {author} {\bibfnamefont {R.}~\bibnamefont {Verresen}}, \bibinfo {author} {\bibfnamefont {M.}~\bibnamefont {Knap}},\ and\ \bibinfo {author} {\bibfnamefont {F.}~\bibnamefont {Pollmann}},\ }\href {https://doi.org/10.1103/PhysRevX.10.011047} {\bibfield  {journal} {\bibinfo  {journal} {Phys. Rev. X}\ }\textbf {\bibinfo {volume} {10}},\ \bibinfo {pages} {011047} (\bibinfo {year} {2020})}\BibitemShut {NoStop}%
\bibitem [{\citenamefont {Rakovszky}\ \emph {et~al.}(2020)\citenamefont {Rakovszky}, \citenamefont {Sala}, \citenamefont {Verresen}, \citenamefont {Knap},\ and\ \citenamefont {Pollmann}}]{rakovszkyStatisticalLocalizationStrong2020}%
  \BibitemOpen
  \bibfield  {author} {\bibinfo {author} {\bibfnamefont {T.}~\bibnamefont {Rakovszky}}, \bibinfo {author} {\bibfnamefont {P.}~\bibnamefont {Sala}}, \bibinfo {author} {\bibfnamefont {R.}~\bibnamefont {Verresen}}, \bibinfo {author} {\bibfnamefont {M.}~\bibnamefont {Knap}},\ and\ \bibinfo {author} {\bibfnamefont {F.}~\bibnamefont {Pollmann}},\ }\href {https://doi.org/10.1103/PhysRevB.101.125126} {\bibfield  {journal} {\bibinfo  {journal} {Phys. Rev. B}\ }\textbf {\bibinfo {volume} {101}},\ \bibinfo {pages} {125126} (\bibinfo {year} {2020})}\BibitemShut {NoStop}%
\bibitem [{\citenamefont {Yoshinaga}\ \emph {et~al.}(2022)\citenamefont {Yoshinaga}, \citenamefont {Hakoshima}, \citenamefont {Imoto}, \citenamefont {Matsuzaki},\ and\ \citenamefont {Hamazaki}}]{yoshinagaEmergenceHilbertSpace2022}%
  \BibitemOpen
  \bibfield  {author} {\bibinfo {author} {\bibfnamefont {A.}~\bibnamefont {Yoshinaga}}, \bibinfo {author} {\bibfnamefont {H.}~\bibnamefont {Hakoshima}}, \bibinfo {author} {\bibfnamefont {T.}~\bibnamefont {Imoto}}, \bibinfo {author} {\bibfnamefont {Y.}~\bibnamefont {Matsuzaki}},\ and\ \bibinfo {author} {\bibfnamefont {R.}~\bibnamefont {Hamazaki}},\ }\href {https://doi.org/10.48550/arXiv.2111.05586} {\bibinfo {title} {Emergence of {{Hilbert Space Fragmentation}} in {{Ising Models}} with a {{Weak Transverse Field}}}} (\bibinfo {year} {2022}),\ \Eprint {https://arxiv.org/abs/2111.05586} {arXiv:2111.05586 [cond-mat, physics:quant-ph]} \BibitemShut {NoStop}%
\bibitem [{\citenamefont {Moudgalya}\ \emph {et~al.}(2022{\natexlab{b}})\citenamefont {Moudgalya}, \citenamefont {Bernevig},\ and\ \citenamefont {Regnault}}]{moudgalyaQuantumManyBodyScars2022a}%
  \BibitemOpen
  \bibfield  {author} {\bibinfo {author} {\bibfnamefont {S.}~\bibnamefont {Moudgalya}}, \bibinfo {author} {\bibfnamefont {B.}~\bibnamefont {Bernevig}},\ and\ \bibinfo {author} {\bibfnamefont {N.}~\bibnamefont {Regnault}},\ }\bibfield  {journal} {\bibinfo  {journal} {Rep. Prog. Phys.}\ }\textbf {\bibinfo {volume} {85}},\ \href {https://doi.org/10.1088/1361-6633/ac73a0} {10.1088/1361-6633/ac73a0} (\bibinfo {year} {2022}{\natexlab{b}})\BibitemShut {NoStop}%
\bibitem [{\citenamefont {Khemani}\ \emph {et~al.}(2020)\citenamefont {Khemani}, \citenamefont {Hermele},\ and\ \citenamefont {Nandkishore}}]{khemaniLocalizationHilbertSpace2020}%
  \BibitemOpen
  \bibfield  {author} {\bibinfo {author} {\bibfnamefont {V.}~\bibnamefont {Khemani}}, \bibinfo {author} {\bibfnamefont {M.}~\bibnamefont {Hermele}},\ and\ \bibinfo {author} {\bibfnamefont {R.}~\bibnamefont {Nandkishore}},\ }\href {https://doi.org/10.1103/PhysRevB.101.174204} {\bibfield  {journal} {\bibinfo  {journal} {Phys. Rev. B}\ }\textbf {\bibinfo {volume} {101}},\ \bibinfo {pages} {174204} (\bibinfo {year} {2020})}\BibitemShut {NoStop}%
\bibitem [{\citenamefont {Moudgalya}\ \emph {et~al.}(2020{\natexlab{b}})\citenamefont {Moudgalya}, \citenamefont {Bernevig},\ and\ \citenamefont {Regnault}}]{moudgalyaQuantumManybodyScars2020}%
  \BibitemOpen
  \bibfield  {author} {\bibinfo {author} {\bibfnamefont {S.}~\bibnamefont {Moudgalya}}, \bibinfo {author} {\bibfnamefont {B.~A.}\ \bibnamefont {Bernevig}},\ and\ \bibinfo {author} {\bibfnamefont {N.}~\bibnamefont {Regnault}},\ }\href {https://doi.org/10.1103/PhysRevB.102.195150} {\bibfield  {journal} {\bibinfo  {journal} {Phys. Rev. B}\ }\textbf {\bibinfo {volume} {102}},\ \bibinfo {pages} {195150} (\bibinfo {year} {2020}{\natexlab{b}})}\BibitemShut {NoStop}%
\bibitem [{\citenamefont {Martyn}\ \emph {et~al.}(2023)\citenamefont {Martyn}, \citenamefont {Liu}, \citenamefont {Chin},\ and\ \citenamefont {Chuang}}]{martynEfficientFullycoherentQuantum2023}%
  \BibitemOpen
  \bibfield  {author} {\bibinfo {author} {\bibfnamefont {J.~M.}\ \bibnamefont {Martyn}}, \bibinfo {author} {\bibfnamefont {Y.}~\bibnamefont {Liu}}, \bibinfo {author} {\bibfnamefont {Z.~E.}\ \bibnamefont {Chin}},\ and\ \bibinfo {author} {\bibfnamefont {I.~L.}\ \bibnamefont {Chuang}},\ }\href {https://doi.org/10.1063/5.0124385} {\bibfield  {journal} {\bibinfo  {journal} {J. Chem. Phys.}\ }\textbf {\bibinfo {volume} {158}},\ \bibinfo {pages} {024106} (\bibinfo {year} {2023})}\BibitemShut {NoStop}%
\bibitem [{\citenamefont {Yamamoto}\ and\ \citenamefont {Yoshioka}(2024)}]{yamamotoRobustAngleFinding2024}%
  \BibitemOpen
  \bibfield  {author} {\bibinfo {author} {\bibfnamefont {S.}~\bibnamefont {Yamamoto}}\ and\ \bibinfo {author} {\bibfnamefont {N.}~\bibnamefont {Yoshioka}},\ }\href {https://doi.org/10.48550/arXiv.2402.03016} {\bibinfo {title} {Robust {{Angle Finding}} for {{Generalized Quantum Signal Processing}}}} (\bibinfo {year} {2024}),\ \Eprint {https://arxiv.org/abs/2402.03016} {arXiv:2402.03016 [quant-ph]} \BibitemShut {NoStop}%
\bibitem [{\citenamefont {Laneve}\ and\ \citenamefont {Wolf}(2025)}]{laneveMultivariatePolynomialsAchievable2025a}%
  \BibitemOpen
  \bibfield  {author} {\bibinfo {author} {\bibfnamefont {L.}~\bibnamefont {Laneve}}\ and\ \bibinfo {author} {\bibfnamefont {S.}~\bibnamefont {Wolf}},\ }\href {https://doi.org/10.22331/q-2025-02-20-1641} {\bibfield  {journal} {\bibinfo  {journal} {Quantum}\ }\textbf {\bibinfo {volume} {9}},\ \bibinfo {pages} {1641} (\bibinfo {year} {2025})}\BibitemShut {NoStop}%
\bibitem [{\citenamefont {Schulz}\ \emph {et~al.}(2019)\citenamefont {Schulz}, \citenamefont {Hooley}, \citenamefont {Moessner},\ and\ \citenamefont {Pollmann}}]{schulzStarkManyBodyLocalization2019}%
  \BibitemOpen
  \bibfield  {author} {\bibinfo {author} {\bibfnamefont {M.}~\bibnamefont {Schulz}}, \bibinfo {author} {\bibfnamefont {C.~A.}\ \bibnamefont {Hooley}}, \bibinfo {author} {\bibfnamefont {R.}~\bibnamefont {Moessner}},\ and\ \bibinfo {author} {\bibfnamefont {F.}~\bibnamefont {Pollmann}},\ }\href {https://doi.org/10.1103/PhysRevLett.122.040606} {\bibfield  {journal} {\bibinfo  {journal} {Phys. Rev. Lett.}\ }\textbf {\bibinfo {volume} {122}},\ \bibinfo {pages} {040606} (\bibinfo {year} {2019})}\BibitemShut {NoStop}%
\bibitem [{\citenamefont {{van Nieuwenburg}}\ \emph {et~al.}(2019)\citenamefont {{van Nieuwenburg}}, \citenamefont {Baum},\ and\ \citenamefont {Refael}}]{vannieuwenburgBlochOscillationsManybody2019}%
  \BibitemOpen
  \bibfield  {author} {\bibinfo {author} {\bibfnamefont {E.}~\bibnamefont {{van Nieuwenburg}}}, \bibinfo {author} {\bibfnamefont {Y.}~\bibnamefont {Baum}},\ and\ \bibinfo {author} {\bibfnamefont {G.}~\bibnamefont {Refael}},\ }\href {https://doi.org/10.1073/pnas.1819316116} {\bibfield  {journal} {\bibinfo  {journal} {Proc. Natl. Acad. Sci. U.S.A.}\ }\textbf {\bibinfo {volume} {116}},\ \bibinfo {pages} {9269} (\bibinfo {year} {2019})}\BibitemShut {NoStop}%
\bibitem [{\citenamefont {Boidi}\ \emph {et~al.}(2025)\citenamefont {Boidi}, \citenamefont {Aharony}, \citenamefont {{Entin-Wohlman}}, \citenamefont {Hallberg},\ and\ \citenamefont {Proetto}}]{boidiCrossoverWannierStarkLocalization2025}%
  \BibitemOpen
  \bibfield  {author} {\bibinfo {author} {\bibfnamefont {N.~A.}\ \bibnamefont {Boidi}}, \bibinfo {author} {\bibfnamefont {A.}~\bibnamefont {Aharony}}, \bibinfo {author} {\bibfnamefont {O.}~\bibnamefont {{Entin-Wohlman}}}, \bibinfo {author} {\bibfnamefont {K.}~\bibnamefont {Hallberg}},\ and\ \bibinfo {author} {\bibfnamefont {C.}~\bibnamefont {Proetto}},\ }\href {https://doi.org/10.48550/arXiv.2502.04866} {\bibinfo {title} {Crossover from {{Wannier-Stark}} localization to charge density waves for interacting spinless fermions in one dimension}} (\bibinfo {year} {2025}),\ \Eprint {https://arxiv.org/abs/2502.04866} {arXiv:2502.04866 [cond-mat]} \BibitemShut {NoStop}%
\bibitem [{\citenamefont {Busch}\ and\ \citenamefont {Penson}(1987)}]{buschTightbindingElectronsOpen1987}%
  \BibitemOpen
  \bibfield  {author} {\bibinfo {author} {\bibfnamefont {U.}~\bibnamefont {Busch}}\ and\ \bibinfo {author} {\bibfnamefont {K.~A.}\ \bibnamefont {Penson}},\ }\href {https://doi.org/10.1103/PhysRevB.36.9271} {\bibfield  {journal} {\bibinfo  {journal} {Phys. Rev. B}\ }\textbf {\bibinfo {volume} {36}},\ \bibinfo {pages} {9271} (\bibinfo {year} {1987})}\BibitemShut {NoStop}%
\bibitem [{\citenamefont {Vandersypen}\ and\ \citenamefont {Chuang}(2005)}]{vandersypenNMRTechniquesQuantum2005}%
  \BibitemOpen
  \bibfield  {author} {\bibinfo {author} {\bibfnamefont {L.~M.~K.}\ \bibnamefont {Vandersypen}}\ and\ \bibinfo {author} {\bibfnamefont {I.~L.}\ \bibnamefont {Chuang}},\ }\href {https://doi.org/10.1103/RevModPhys.76.1037} {\bibfield  {journal} {\bibinfo  {journal} {Rev. Mod. Phys.}\ }\textbf {\bibinfo {volume} {76}},\ \bibinfo {pages} {1037} (\bibinfo {year} {2005})}\BibitemShut {NoStop}%
\bibitem [{\citenamefont {Weinberg}\ and\ \citenamefont {Bukov}(2017)}]{weinbergQuSpinPythonPackage2017}%
  \BibitemOpen
  \bibfield  {author} {\bibinfo {author} {\bibfnamefont {P.}~\bibnamefont {Weinberg}}\ and\ \bibinfo {author} {\bibfnamefont {M.}~\bibnamefont {Bukov}},\ }\href {https://doi.org/10.21468/SciPostPhys.2.1.003} {\bibfield  {journal} {\bibinfo  {journal} {SciPost Physics}\ }\textbf {\bibinfo {volume} {2}},\ \bibinfo {pages} {003} (\bibinfo {year} {2017})}\BibitemShut {NoStop}%
\bibitem [{\citenamefont {Weinberg}\ and\ \citenamefont {Bukov}(2019)}]{weinbergQuSpinPythonPackage2019}%
  \BibitemOpen
  \bibfield  {author} {\bibinfo {author} {\bibfnamefont {P.}~\bibnamefont {Weinberg}}\ and\ \bibinfo {author} {\bibfnamefont {M.}~\bibnamefont {Bukov}},\ }\href {https://doi.org/10.21468/SciPostPhys.7.2.020} {\bibfield  {journal} {\bibinfo  {journal} {SciPost Physics}\ }\textbf {\bibinfo {volume} {7}},\ \bibinfo {pages} {020} (\bibinfo {year} {2019})}\BibitemShut {NoStop}%
\bibitem [{\citenamefont {Iadecola}(2025)}]{iadecolaSymmetryFragmentation2025}%
  \BibitemOpen
  \bibfield  {author} {\bibinfo {author} {\bibfnamefont {T.}~\bibnamefont {Iadecola}},\ }\href {https://doi.org/10.48550/arXiv.2510.06333} {\bibinfo {title} {Symmetry {{Fragmentation}}}} (\bibinfo {year} {2025}),\ \Eprint {https://arxiv.org/abs/2510.06333} {arXiv:2510.06333 [quant-ph]} \BibitemShut {NoStop}%
\bibitem [{\citenamefont {Motlagh}\ and\ \citenamefont {Wiebe}(2024)}]{motlaghGeneralizedQuantumSignal2024}%
  \BibitemOpen
  \bibfield  {author} {\bibinfo {author} {\bibfnamefont {D.}~\bibnamefont {Motlagh}}\ and\ \bibinfo {author} {\bibfnamefont {N.}~\bibnamefont {Wiebe}},\ }\href {https://doi.org/10.1103/PRXQuantum.5.020368} {\bibfield  {journal} {\bibinfo  {journal} {PRX Quantum}\ }\textbf {\bibinfo {volume} {5}},\ \bibinfo {pages} {020368} (\bibinfo {year} {2024})}\BibitemShut {NoStop}%
\end{thebibliography}%

\end{document}